\RequirePackage{fix-cm}
\documentclass[twocolumn]{svjour3}       
\smartqed  
\usepackage{graphicx}
\usepackage [latin1]{inputenc}
\usepackage{marvosym}
\usepackage{amssymb}
\usepackage{indentfirst}
\usepackage{graphicx}
\usepackage{epstopdf}
\usepackage{float}
\usepackage{subfig}
\usepackage{amssymb}
\usepackage{stmaryrd}
\usepackage{amsfonts}
\usepackage{amsmath}
\usepackage{latexsym}
\usepackage{color}
\usepackage{CJK}
\usepackage{mathrsfs}
\usepackage{algorithm} 
\usepackage{algorithmic} 
\usepackage{multirow} 
\usepackage{amsmath}
\usepackage{xcolor}
\usepackage{graphicx}
\usepackage{mathptmx}
\usepackage{indentfirst}
\usepackage{booktabs}
\usepackage{float}
\usepackage{subfig}
\usepackage{amssymb}
\usepackage{stmaryrd}
\usepackage{dsfont}
\usepackage{amssymb}
\usepackage{amsfonts}
\usepackage{amsmath}
\usepackage{latexsym}
\usepackage{color}
\usepackage{cite}
\usepackage{mathrsfs}
\usepackage{epstopdf}
\journalname{Nonlinear Dyn}
\begin{document}

\title{A fractional-order SEIHDR model for COVID-19 with inter-city networked coupling effects\thanks{This work is supported by the Natural Science Foundation of Beijing Municipality (No. Z180005) and the National
Nature Science Foundation of China (No. 61772063).
}}


\author{Zhenzhen Lu        \and
        Yongguang Yu       \and
        YangQuan Chen      \and
        Guojian Ren        \and
        Conghui Xu         \and
        Shuhui Wang        \and
        Zhe Yin
}


\institute{Z. Lu \and Y. Yu ${(\textrm{\Letter})}$ \and G. Ren \and C. Xu \and S. Wang \and Z. Yin\at
              Department of Mathematics, Beijing Jiaotong University, Beijing, 100044, P.R.China \\
              \email{ygyu@bjtu.edu.cn}           
           \and
           Y. Chen ${(\textrm{\Letter})}$ \at
              Mechatronics, Embedded Systems and Automation Lab, University of California, Merced, CA 95343, USA\\
              \email{ychen53@ucmerced.edu}
}
\date{Received: date / Accepted: date}

\maketitle

\begin{abstract}
A novel coronavirus, designated as COVID-19, emerged in Wuhan, China, at the end of 2019. In this paper, a mathematical model is proposed to analyze the dynamic behavior of COVID-19. Based on inter-city networked coupling effects, a fractional-order SEIHDR system with the real-data from 23 January to 18 March, 2020 of COVID-19 is discussed. Meanwhile, hospitalized individuals and the mortality rates of three types of individuals (exposed, infected and hospitalized) are firstly taken into account in the proposed model. And infectivity of individuals during incubation is also considered in this paper. By applying least squares method and predictor-correctors scheme, the numerical solutions of the proposed system in the absence of the inter-city network and with the inter-city network are stimulated by using the real-data from 23 January to $18-m$ March, 2020 where $m$ is equal to the number of prediction days. Compared with integer-order system ($\alpha=0$), the fractional-order model without network is validated to have a better fitting of the data on Beijing, Shanghai, Wuhan, Huanggang and other cities. In contrast to the case without network, the results indicate that the inter-city network system may be not a significant case to virus spreading for China because of the lock down and quarantine measures, however, it may have an impact on cities that have not adopted city closure. Meanwhile, the proposed model better fits the data from 24 February to 31, March in Italy, and the peak number of confirmed people is also predicted by this fraction-order model. Furthermore, the existence and uniqueness of a bounded solution under the initial condition are considered in the proposed system. Afterwards, the basic reproduction number $R_0$ is analyzed and it is found to hold a threshold: the disease-free equilibrium point is locally asymptotically stable when $R_0\le 1$, which provides a theoretical basis for whether COVID-19 will become a pandemic in the future.
\end{abstract}

\section{Introduction}
\label{intro}
In December 2019, the first case of respiratory disease caused by a novel coronavirus was identified in Wuhan City, Hubei Province, China. Contrary to the initial observations \cite{Cheng_2020}, COVID-19 does spread from person to person as confirmed in \cite{Chan_2020}. It quickly spread to all parts of the country and parts of Southeast Asia, North America and Europe. As of 5 February, 2020, more than $24550$ cases of coronavirus disease 2019 (COVID-19) had been confirmed, including over $190$ cases outside of China, and more than $490$ reported deaths globally. Several intervention strategies have been implemented in China in order to contain the epidemic. Whereas assessing the intervention measures of COVID-19 epidemic poses a major health concern.

Wuhan is the capital of Hubei Province, as the major air and train transportation hub of central China with huge population movement during the period of spring festival.  The geographical factors plus self-sustaining human-to- human transmission in the community of the new coronavirus have added radical difficulties to the prevention and interference of this epidemic. The outflow of domestic passenger volumes from Wuhan was estimated to be $5$ million during the spring festival, about $0.7$ of which went to other cities in Hubei province.  which made it difficult for the infected people to track back to the population. Government has progressively implemented metropolitan-wide quarantine of Wuhan since 23-24 January, 2020 which ultimately contributes a lot of the containment of virus. The epidemic was seeded by a force of frequent human contact generating from communication. Prasse et al. proposed a network-based discrete model to describe this condition and they found the network-based modelling was beneficial for an accurate forecast of the epidemic outbreak \cite{prasse2020network}. Peng et al. constructed and analyzed a generalized SEIR model to analyze COVID-19 and their work showed that the outbreak of COVID-19 in China could be dated back to the end of December 2019 \cite{Peng_2020}. In addition, the basic reproduction number $R_0$ is 'the expected number of secondary cases produced, in a completely susceptible population, by a typical infective individual'. If $R_0 \le 1$, on average an infected individual produces less than one new infected individual over the course of its infectious period, and the infection cannot grow. Conversely, if $R_0 > 1$, each infected individual produces, on average, more than one new infection, and the disease can invade the population \cite{van_den_Driessche_2002}. There have been abundant articles estimating the basic reproduction number $R_0$ of COVID-19 to determine whether the epidemic disease will spread widely \cite{Liu_2020,Lai_2020,Sahafizadeh_2020}. Meanwhile, in the process of infectious disease transmission, susceptible individuals contact infected individuals and become infected with a certain probability. There is a lot of evidence that the incidence rate is an important tool to describe this process \cite{anderson1992infectious,bailey1975mathematical,upadhyay2019dynamics}. It represents the infection ability of a infected individuals in per-unit time. So because of high infectiousness of COVID-19, the bilinear incidence rate is always considered. And the classical SEIR model assumes that the incubation of infected person is not infectious, but this assumption is quite different from the infection characteristics of the new coronavirus infection COVID-19 \cite{Tang_2020}. What's more, due to a variety of special conditions, patients with COVID-19 cannot be admitted to hospital for immediate treatment, and individuals infected with COVID-19 have a high mortality rate both in the incubation period and no-treatment period.

It is worthing to point out that the state of epidemic models does not depend on its history in the classical integer-order epidemic model. However, in real life, the spread of infectious diseases depends not only on its current state, but also on its past state. Smethurst et al. find that the waiting times between doctor visits for a patient follow a power law model  \cite{Smethurst_2001}. At the same time, a power law waiting time distribution $P[J_n > t] = Bx^{-\alpha}$ leads to Caputo fractional time derivative ${}_{t_0}^CD_t^\alpha$ of the same order \cite{2011}. And the Caputo fractional-order derivatives allows traditional initial and boundary conditions when dealing with real-world problems. Not only that, its derivative for a constant is zero. What's more, the precision of the Caputo fractional-order can supersede the integer order resulting from its changes at every instant of time and nonlocal behavior. Khan et al. considered a fractional-order model to describe the brief details of interactional among the bats and unknown hosts, then among the people and the infections reservoir (seafood market) and found the fractional model can be helpful for the infection minimization \cite{Khan_2020}. Yu et al. establish a novel fractional time delay dynamic system (FTDD) to describe the local outbreak of COVID-19 and  the reconstructed coefficients are used to predict the trend of the Corona-Virus \cite{chen2020the}. Shaikh et al. estimated the effectiveness of preventive measures and various mitigations, predicting future outbreaks and potential control strategies using a Bats-Hosts-Reservoir-People transmission fractional-order COVID-19 model \cite{shaikh2020mathematical}. Xu et al. proposed a generalized fractional-order SEIQRD model and this model has a basis guiding significance for the prediction of the possible outbreak of infectious diseases like COVID-19 and other insect diseases in the future \cite{xu_2020}.

The purpose of this paper is to incorporate time-fractional-order and inter-city networked coupling effects into a SEIHDR epidemic model to investigate the dynamic behavior of COVID-19. There are a lot of research for COVID-19 that
not only individuals are contagious, but those in the incubation period may have the same incidence rate as well as those with symptoms. So the infectiousness of the infected person's incubation period is considered in this paper. Also, considering an individual hospitalized with COVID-19, the different mortality rates are investigated in this model. Thus based on the above analysis and the memorability of Caputo fractional-order derivatives, a fractional-order SEIHDR model for COVID-19 with inter-city networked coupling effects are built. Should such a model can be fitted with data reasonably well, some epidemic parameters can be extracted. Then, based on the official data given by NHC (National Health Commission of the people's Republic of China \cite{LI_2020}) every day, several numerical examples are exhibited to verify the rationality of the fractional-order model and effectiveness of inter-city networked coupling. Compared with integer-order system ($\alpha=0$), the fractional-order model without network is validated to have a better fitting of the data on Beijing, Shanghai, Wuhan, Huanggang and other cities. In the case of inter-city networked coupling effect, the results indicate that this system may be not a significant case to virus spreading for China, but may have an impact on preventive measures in other countries. Finally, the stability of disease-free equilibrium point is obtained by the basic reproduction number $R_0$, which provide theoretical significance for the development of COVID-19 in futher.

The rest of the paper is organized as follows. In Section $2$, a fractional SEIHDR epidemic model for COVID-19 is
formulated. In Section $3$, some dynamical behaviors of the proposed system are analyzed. In Section $4$, some numerical simulations are presented to illustrate theoretical results according to the official dates. Finally, a brief discussion is given in Section $5$.

\section{Model Development}

Fractional-order calculus have been found wide applications to model dynamics processes in many well-known fields of science, engineering, biology, medicine, and many other \cite{Huo_2015,Kheiri_2019,Almeida_2018,Owolabi_2019,Sierociuk_2015}. Before presenting the epidemic model in fractional derivative, some necessary preliminaries are introduced.

\subsection{Preliminaries}
\label{sec:1}
In this subsection, some definitions and results are introduced firstly.

\textbf{Definition 2.1. \cite{Zwillinger_1992}} A Gamma function of  $\alpha>0 $ is defined by
\[\Gamma(\alpha)= \int_{0}^\infty{x^{\alpha-1}e^{-x}dx}.\]

\textbf{Definition 2.2. \cite{Zwillinger_1992}} For any $t > t_0$, the time Caputo fractional derivative of order $\alpha $ ($n< \alpha< n+1$) with the lower limit $t_0\ge 0$ for a function $f( t)\in \mathbb{R}$ is defined by
\[{}_{t_0}^CD_t^\alpha f\left( t \right)=\frac{{{d ^\alpha }f\left( {t} \right)}}{{d {t^\alpha }}} = \frac{1}{{\Gamma
\left( {1 - \alpha } \right)}}\int_{t_0}^t {\frac{{d^n f\left( {s} \right)}}{{d^n s}}} \frac{{ds}}{{{{\left( {t - s}
\right)}^{\alpha+n-1} }}},\]
where $\Gamma\left( \cdot \right)$ is the Gamma function.

\textbf{Remark 2.1. } When $\alpha=n$, \[{}_{t_0}^CD_t^\alpha f\left( t \right)=f^{(n)}(t).\]

\textbf{Definition 2.3. \cite{Li_2009} } A constant ${x^*}$ is an equilibrium point of the Caputo fractional dynamical system:
\[{}_{{t_0}}^CD_t^\alpha x(t) = f(t,x),\quad x({t_0}) \ge 0,\]
if and only if $f(t,{x^*}) = 0$.

\textbf{Lemma 2.1. \cite{Li_2010}} Consider the fractional-order system:
\[{}_{{t_0}}^CD_t^\alpha x(t) = f(t,x),~{t_0} > 0,\]
with the initial condition $x({t_0}) = {x_0}$, where $\alpha  \in (0,1]$ and $f:[{t_0},\infty ) \times \Gamma  \to\mathbb{R}^n,\Gamma  \in \mathbb{R}^n$, if $f(t,x)$ satisfies the local Lipschitz condition with respect to $x$, there exists a unique solution of the above system.

\textbf{Lemma 2.2. \cite{article}} Consider the following fractional-order system:
\[{}_{{t_0}}^CD_t^\alpha x(t) = f(x),\quad {x_{0}} = x({t_0}) > 0,\]
with $0 < \alpha  \le 1$, $x \in R$. The equilibrium points of the above system are calculated by solving the
following equation: $f(x)=0$. These points are locally asymptotically stable if and only if all eigenvalues $\lambda$
of the Jacobian matrix $J = \frac{{\partial f}}{{\partial x}}$ at $x$ evaluated of the equilibrium points satisfy
$|arg(\lambda )| >\frac{{\alpha \pi }}{2}$.

\subsection{System description}
\label{sec:2}
A novel coronavirus disease, named COVID-19, broke out in Wuhan, Hubei province, China, in December 2019. A great deal of epidemic models exists to describe spread of infectious diseases mathematically \cite{1}. However, the outbreak of COVID-19 began during the period of spring festival in China, which large population movement made cities more interconnected. Meanwhile, Wuhan is the capital of Hubei Province, as the major air and train transportation hub of central China, which makes it necessary to built a inter-city networked to analyze the spread of COVID-19. And due to limited medical conditions, individuals in incubation and infection periods cannot be hospitalized immediately for treatment, which increases the mortality of COVID-19. Moreover, Tang et al proposed that the incubation of infected person of COVID-19 is infectious \cite{Tang_2020}. Based on the above analysis, a fractional-order SEIHDR model with inter-city networked coupling effects in this paper is developed as follows:
\begin{equation}
\begin{aligned}
\left\{ \begin{array}{l}
 {}_{{0}}^CD_t^\alpha S_k =  - \sum\limits_{j = 1}^n {\beta _{kj} (\frac{S_k I_j}{N_k}+\frac{S_k E_j}{N_k})},  \\
 {}_{{0}}^CD_t^\alpha E_k= \sum\limits_{j = 1}^n {\beta _{kj} (\frac{S_k I_j}{N_k}+\frac{S_k E_j}{N_k})} - \mu _{1k} E_k  - r_k E_k , \\
 {}_{{0}}^CD_t^\alpha I_k={r_k}{E_k}-{\delta_k}{I_k}-\mu _{2k} I_k,\\
{}_{{0}}^CD_t^\alpha H_k  = {\delta_k}{I_k}  - {\lambda_k}(t){H_k}  -  \kappa _k H_k , \\
 {}_{{0}}^CD_t^\alpha R_k ={\lambda_k}{H_k},  \\
 {}_{{0}}^CD_t^\alpha D_k=\mu _{1k} E_k +\mu _{2k} I_k+\kappa_k H_k,
\end{array} \right.
\end{aligned}
\end{equation}
with the initial condition
\begin{equation}
\begin{aligned}
&S_{\rm{k}} (0) = S_{\rm{k}} (0) \ge 0,~~E_{\rm{k}} (0) = E_{\rm{k}} (0) \ge 0,\\
&I_{\rm{k}} (0) = I_{\rm{k}} (0) \ge 0,~~H_{\rm{k}} (0) = H_{\rm{k}} (0) \ge 0,\\
&R_{\rm{k}} (0) = R_{\rm{k}} (0) \ge 0,~~D_{\rm{k}} (0) = D_{\rm{k}} (0) \ge 0.
\end{aligned}
\end{equation}
The total population of city $k$ is denoted by $N_k$ which is further classified into six subgroups where $S_k(t)$, $E_k(t)$, $I_k(t)$, $H_k(t)$, $D_k(t)$ and $R_k(t)$ $(k=1,2,...,n)$ present the number of the susceptible, exposed, infective (infected but not hospitalized), hospitalization, death and recovered individuals at time $t$ and city $k$, respectively. The susceptible people $S_k$ infected after the interaction with $E_j$ and $I_j$, given by $\sum\limits_{j = 1}^n {\beta _{kj} (\frac{S_k I_j}{N_k}+\frac{S_k E_j}{N_k})}$, where $\beta_{kj}$ $(k,j=1,2,...,n)$ is the disease transmission coefficient, respectively. The death people $D_k$ includes death during exposure $\mu _{1k} E_k$, infection $\mu _{2k}I_k$, and hospitalization $\kappa_k(t) H_k$, where $\mu_{ik}$ $(i=1,2)$ and $\kappa_k(t)$ implies the disease-related death rate. The parameters $\lambda_k(t)$ be the recovery rate through hospitalization; $r_k$ denotes the transit rate of the exposed individuals $E_k$; $\delta_k$ be hospitalization rate of the infective individuals; $t=0$ represents 23 January, 2020. Furthermore, $\beta_{kj}$, $\mu_{ik}$, $\delta_k$ and $r_k$ $(i=1,2)$ are positive constants; $\lambda_k(t)$ and $\kappa_k(t)$ are bounded function, i.e. $|\lambda_k(t)|\le M_{1k}$ and $|\kappa_k(t)|\le M_{2k}$ for all $t\ge 0$, where $M_{1k}$ and $M_{2k}$ are positive constants.

\section{Model Analysis}

In this section, some dynamical behaviors of the proposed system (1) are analyzed.

\subsection{Nonnegativity and  Boundedness}
\label{sec:1}
It is significant to demonstrate the existence, uniqueness and boundedness of a nonnegative solutions for system (1) before implementing its numerical process. Thus this subsection moves to the discussion of proprieties mentioned above.

\textbf{Theorem 3.1.} For any given initial condition $(S_k(0),E_k(0)$, $I_k(0), H_k(0),R_k(0),D_k(0))\ge (\not\equiv) (0,0,0,0,0,0) $, there exists a unique nonnegative and boundedenss solution $(S_k(t),E_k(t)$, $I_k(t),H_k(t),R_k(t),D_k(t))$ of system (1) for all $k=1,2,...,n$ and $t>0$ where $t=0$ represents 23 January, 2020.

\textbf{Proof:} Let $N_k=S_k+E_k+I_k+H_k+R_k+D_k$. Adding all equations of system (1) gives ${}_{{0}}^CD_t^\alpha N_k=0$ which $N_k=M_k$ is a positive constant. So one has $S_k\le M_k$, $E_k\le M_k$, $I_k\le M_k$, $H_k\le M_k$, $R_k\le M_k$ and $D_k\le M_k$ where $k=1,2,...,n$.\\
Let $X_k=(S_k,E_k,I_k,H_k,R_k,D_k)$, $\overline X=(\overline {S}_k,\overline {E}_k,\overline {I}_k,\overline {H}_k,\overline {R}_k,\overline {D}_k)$ and $F_k=(f_{1k},f_{2k},f_{3k},f_{4k},f_{5k},f_{6k})$ where
\begin{equation}
\begin{aligned}
&f_{1k}=- \sum\limits_{j = 1}^n {\beta _{kj} (\frac{S_k I_j}{N_k}+\frac{S_k E_j}{N_k})} ,\\
&f_{2k}=\sum\limits_{j = 1}^n {\beta _{kj} (\frac{S_k I_j}{N_k}+\frac{S_k E_j}{N_k})} - \mu _{1k} E_k  - r_k E_k,\\
&f_{3k}={r_k}{E_k}-{\delta_k}{I_k}-\mu _{2k} I_k,\\
&f_{4k}={\delta_k}{I_k}  - {\lambda_k}(t){H_k}  -  \kappa _k H_k ,\\
&f_{5k}={\lambda_k}{H_k},\\
&f_{6k}=\mu _{1k} E_k +\mu _{2k} I_k+\kappa_k H_k.
\end{aligned}
\end{equation}
Obviously, one has
\begin{equation}
\begin{aligned}
||F_k(X) - F_k(\overline X )|| &\le ||{f_{1k}}(X) - {f_{1k}}(\overline X )|| + ||{f_{2k}}(X) - {f_{2k}}(\overline X )||\\ &+||{f_{3k}}(X) - {f_{3k}}(\overline X )||+ ||{f_{4k}}(X) - {f_{4k}}(\overline X )||\\
&+ ||{f_{5k}}(X) - {f_{5k}}(\overline X )||+ ||{f_{6k}}(X) - {f_{6k}}(\overline X )||\\
&\le L_k||X- \overline X ||,
\end{aligned}
\end{equation}
where $L_k=\max(L_{1k},L_{2k},L_{3k},L_{4k},L_{5k})$, $L_{1k}=\frac{2\beta_{kj}M}{N}$, $L_{2k}=L_{1k}+\mu_{1k}+r_k$, $L_{3k}=r_k+\mu_{2k}+\delta_k$, $L_{4k}=M_{1k}+M_{2k}$ and $L_{5k}=\mu_{1k}+\mu_{2k}+M_{2k}$. So $F_k$ satisfies the local Lipschitz condition with respect to $X_k$. Then there exists a unique boundedenss solution $(S_k(t),E_k(t),I_k(t),H_k(t),R_k(t),D_k(t))_{1\le k\le n}$ of system (1).\\
Furthermore, consider the following auxiliary system:
\begin{equation}
\begin{aligned}
\left\{ \begin{array}{l}
 {}_{{0}}^CD_t^\alpha \underline{S}_k =  - \sum\limits_{j = 1}^n {\beta _{kj} (\frac{\underline{S}_k \underline{I}_j}{N_k}+\frac{\underline{S}_k \underline{E}_j}{N_k})}, \\
 {}_{{0}}^CD_t^\alpha \underline{E}_k= \sum\limits_{j = 1}^n {\beta _{kj} (\frac{\underline{S}_k \underline{I}_j}{N_k}+\frac{\underline{S}_k \underline{E}_j}{N_k})} - \mu _{1k}\underline{E}_k  - r_k \underline{E}_k,  \\
 {}_{{0}}^CD_t^\alpha \underline{I}_k={r_k}\underline{E}_k-{\delta_k}\underline{I}_k-\mu _{2k}\underline{I}_k,\\
{}_{{0}}^CD_t^\alpha \underline{H}_k  = {\delta_k}\underline{I}_k  - {\lambda_k}(t)\underline{H}_k  -  {\kappa _k}(t) \underline{H}_k,  \\
 {}_{{0}}^CD_t^\alpha \underline{R}_k ={\lambda_k}(t)\underline{H}_k,  \\
 {}_{{0}}^CD_t^\alpha \underline{D}_k=\mu _{1k}\underline{E}_k +\mu _{2k} \underline{I}_k+{\kappa_k}(t) \underline{H}_k,\\
\underline {S}_k=\underline {E}_k=\underline {I}_k=\underline {H}_k=\underline {R}_k=\underline {D}_k=0.\nonumber
\end{array} \right.
\end{aligned}
\end{equation}
Through the comparison theorem, it is not difficult to find that $(\underline {S}_k,\underline {E}_k,\underline {I}_k,\underline {H}_k,\underline {R}_k,\underline {D}_k)=(0,0,0,0,0,0)$ is the lower solution of system (1). Thus, one has $S_k\ge 0$, $E_k\ge 0$, $I_k\ge 0$, $H_k\ge 0$, $R_k\ge 0$ and $D_k\ge 0$ for $t\ge 0$ and
$k=1,2,...,n$.\hfill$\square$

\subsection{Stability Analysis}
\label{sec:2}
Turning now to the exploration on local stability of system (1) by considering first the disease-free equilibrium and the basic reproduction number denoted by $R_0$. According to system (1), the disease-free equilibrium for system (1) is $E^0=(S_1(0),0,0,0,0,0,...,S_n(0),0,0,0,0,0)$ where $S_k(0)$ $(k=1,2,...,n)$ are the initial condition of system (1).

\textbf{Theorem 3.2.} The basic reproduction number of system (1) is
\[R_0=\rho ({F_1 (V_{21}  - V_{20} )}{({V_{21} V_{10} })^{-1}}),\]
where $F_1=(\frac{\beta _{kj} S_{_k } (0)}{N_k})$, $V_{10}  = diag(\mu _{11}  + r_1 ,...,\mu _{1n}  + r_n )$,
$V_{20}  = diag( - r_1 ,..., - r_n )$, $V_{21}  = diag(\delta _1  + \mu _{21} ,...,\delta _n  + \mu _{2n} )$ and \\$\rho ({F_1 (V_{21}  - V_{20} )}{({V_{21} V_{10} })^{-1}})$ which is the spectral radius of the matrix $({F_1 (V_{21}  - V_{20} )}{({V_{21} V_{10} })^{-1}})$.

\textbf{Proof:} Let $F_0=(\sum\limits_{j = 1}^n {\frac{\beta _{kj} S_k (I_j  + E_j )}{N_k}})$,
$V_{01}={((\mu_{1k}+r_{k})E_{k})}$, $V_{02}=({\delta_k}{I_k}+\mu_{2k}{I_k}-r_{k}E_k)$, $V_{03}=({\lambda_k}{H_k}+\kappa_{k}{H_k}-\delta_{k}I_k)$. Then taking the derivative of $F_0$, $V_{01}$, $V_{02}$ and $V_{03}$ with respect to $E_k$, $I_k$ and $H_k$ $(k=1,2,...,n)$ at the disease-free equilibrium point $E^0$, one has
\[
F = \left( {\begin{array}{*{20}c}
   {F_1 } & {F_1 } & 0  \\
   0 & 0 & 0  \\
   0 & 0 & 0 \\
\end{array}} \right),
\]
\[V=(V_1,V_2.V_3),\]
where $V_1=(V_{10},0,0),V_2=(V_{20},V_{21},0),V_3=(0,V_{30},V_{31})$, $F_1=(\frac{\beta _{kj} S_{_k } (0)}{N_k})_{1 \le k,j \le n}$, $V_{10}  = diag(\mu _{11}  + r_1 ,....,\mu _{1n}  + r_n )$, $V_{20}  = diag( - r_1 ,...., - r_n )$, $V_{21}  = diag(\delta _1  + \mu _{21} ,....,\delta _n  + \mu _{2n} )$, $V_{30}  = diag(-\delta _1,....,-\delta _n)$ and $V_{31} = diag(\lambda_1 + \kappa_1 ,....,\lambda_n+\kappa_n)$. Then according to the definition of the
basic reproduction number \cite{van_den_Driessche_2002}, one has \[R_0=\rho (FV^{-1})=\rho ({F_1 (V_{21}  - V_{20} )}{({V_{21} V_{10} })^{-1}}).\hfill \square\]

Based on the basic reproduction number $R_0$, the following theorem can be taken into consideration:

\textbf{Theorem 3.3.} If $R_0\le 1$, the disease-free equilibrium point $E^0=(S_1(0),0,0,0,0,0,...,S_n(0),0,0,0,0,0)$ of
system (1) is locally asymptotic stability.

\textbf{Proof:} The disease-free equilibrium point $E^0$ of system (1) is locally asymptotically stable if all eigenvalues of the Jacobian matrix of system (1) at $E^0$
namely,
\[
J^0  = \left( {\begin{array}{*{20}c}
   0 & { - F_1 } & { - F_1 } & 0 & 0 & 0  \\
   0 & {F_1  - V_1 } & {F_1 } & 0 & 0 & 0  \\
   0 & { - V_{20} } & { - V_{21} } & 0 & 0 & 0  \\
   0 & 0 & { - V_{30} } & { - V_{31} } & 0 & 0  \\
   0 & 0 & 0 & { - V_{41} } & 0 & 0  \\
   0 & { - V_{51} } & { - V_{61} } & { - V_{71} } & 0 & 0  \\
\end{array}} \right),
\]
satisfies $|arg(s)|>\frac{\alpha\pi}{2}$ and unstable if for some eigenvalues $s$,
$|arg(s)|<\frac{\alpha\pi}{2}$, where $V_{41}  = diag( \lambda_1(t),...., \lambda_n(t))$, $V_{51}  = diag( \mu_{11},...., \mu_{1n} )$, $V_{61}  = diag( \mu_{21} ,...., \mu_{2n} )$ and $V_{71}  = diag$ $( \kappa_1(t),...., \kappa_n (t))$. One can calculate that the eigenvalues are $s_1=0$ and $s_2=-s(V_{31})$ and
\[
\begin{array}{l}
 s _3  + s_4  = s(F_1  - V_{10}  - V_{21} ), \\
 {s _3}{ s _4}  = s(F_1 (V_{20} - V_{21}) + V_{20}V_{21}), \\
 \end{array}
\]
where $s (V_{31})$, $s(F_1  - V_{10}  - V_{21} )$ and $s(F_1 (V_{20} - V_{21}) + V_{20}V_{21})$ are all eigenvalues of the matrix $V_{31}$, $F_1  - V_{10}  - V_{21}$ and $F_1 (V_{20} - V_{21}) + V_{20}V_{21}$, respectively. Then if $R_0\le 1$, one has $|arg(s)|>\frac{\alpha\pi}{2}$. Thus the disease-free equilibrium point $E^0=(S_1(0),0,0,0,0,0,...,S_n(0),0,0,0,0,0)$ of system (1) is locally asymptotic stability.\hfill$\square$

\textbf{Remark 3.1.} When there is no interaction  between city $k$ and city $j$ (i.e. city $k$ is isolated from other cities), the basic reproduction number in city $k$ is given:
\[R^k_{0}=\frac{\beta_{kk}S_k(0)({\delta_k}+\mu_{2k}+r_k)}{(\delta_k+\mu_{2k})(\mu_{1k}+r_k)N_k}.\]
In this case, the disease-free equilibrium point $E^0_{k}=(S_k(0),0$, $0,0,0,0)$ of city $k$ is local asymptotic stability when the basic number $R^k_{0}\le 1$.

\textbf{Remark 3.2.} It is obvious that the basic number $R^k_{0}$ is dependent from the onset $\lambda_k$ and
$\kappa_k$. The sensitivity of $R^k_{0}$ to the other parameters $\beta_{kk}$, $\mu_{1k}$, $r_k$, $\mu_{2k}$ and
$\delta_k$ are calculated as follows:
\[
A_{\beta _{{\rm{kk}}} } {\rm{ = }}\frac{{\beta _{{\rm{kk}}} }}{{R_0 }}\frac{{\partial R_0 }}{{\partial \beta
_{{\rm{kk}}} }} = 1,~~~A_{\mu _{{\rm{1k}}} } {\rm{ = }}\frac{{\mu _{{\rm{1k}}} }}{{R_0 }}\frac{{\partial R_0
}}{{\partial \mu _{{\rm{1k}}} }} =  - \frac{{\mu _{{\rm{1k}}} }}{{\mu _{{\rm{1k}}}  + r_k }},
\]
\[
A_{\mu _{{\rm{2k}}} } {\rm{ = }}\frac{{\mu _{{\rm{2k}}} }}{{R_0 }}\frac{{\partial R_0 }}{{\partial \mu _{{\rm{2k}}} }}
= -\frac{{r_k }}{{(\delta _k  + \mu _{{\rm{2k}}} )(\delta _k  + \mu _{{\rm{2k}}}  + r_k )}},\]
\[A_{r_{\rm{k}} } {\rm{ =
}}\frac{{r_{\rm{k}} }}{{R_0 }}\frac{{\partial R_0 }}{{\partial r_{\rm{k}} }} = \frac{{r_k (\mu _{{\rm{1k}}}  - \mu
_{{\rm{2k}}}  - \delta _k )}}{{(\delta _k  + \mu _{{\rm{2k}}}  + r_k )}},
\]
\[A_{\delta _{{\rm{k}}} } {\rm{ = }}\frac{{\delta _{{\rm{k}}} }}{{R_0 }}\frac{{\partial R_0 }}{{\partial \mu _{{\rm{2k}}} }}
= -\frac{{r_k }{\delta_k}}{{(\delta _k  + \mu _{{\rm{2k}}} )(\delta _k  + \mu _{{\rm{2k}}}  + r_k )}},
\]
where $A_{\beta_{kk}}$, $A_{\mu_{1k}}$, $A_{\mu_{2k}}$ and $A_{\delta_k}$ denote the normalized sensitivity indexes with respect to $\beta_{kk}$, $\mu_{1k}$, $\mu_{2k}$ and $\delta_k$, respectively. It is remarkable that the $m$ times' increase in $\beta_{kk}$ results in the $m$ times' increase in $R^k_0$, but the $m$ times' increase in $\mu_{1k}$, $\mu_{2k}$, $\delta_k$ and $r_{k}$ results in the $m$ times' decrease in $R^k_0$.

\section{Numerical Analysis}

\begin{figure*}
\centering
\includegraphics[width=0.4\linewidth]{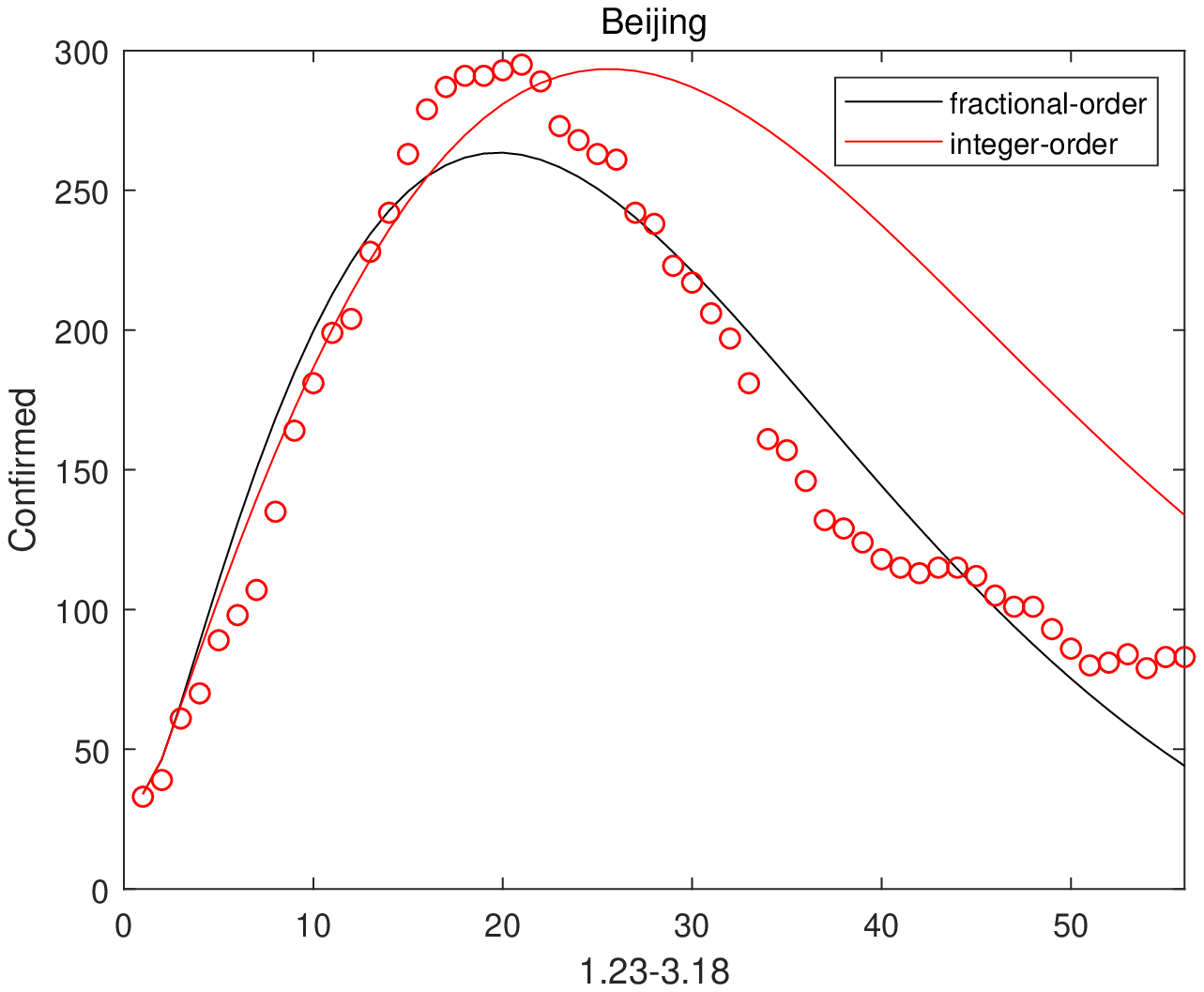}
\includegraphics[width=0.4\linewidth]{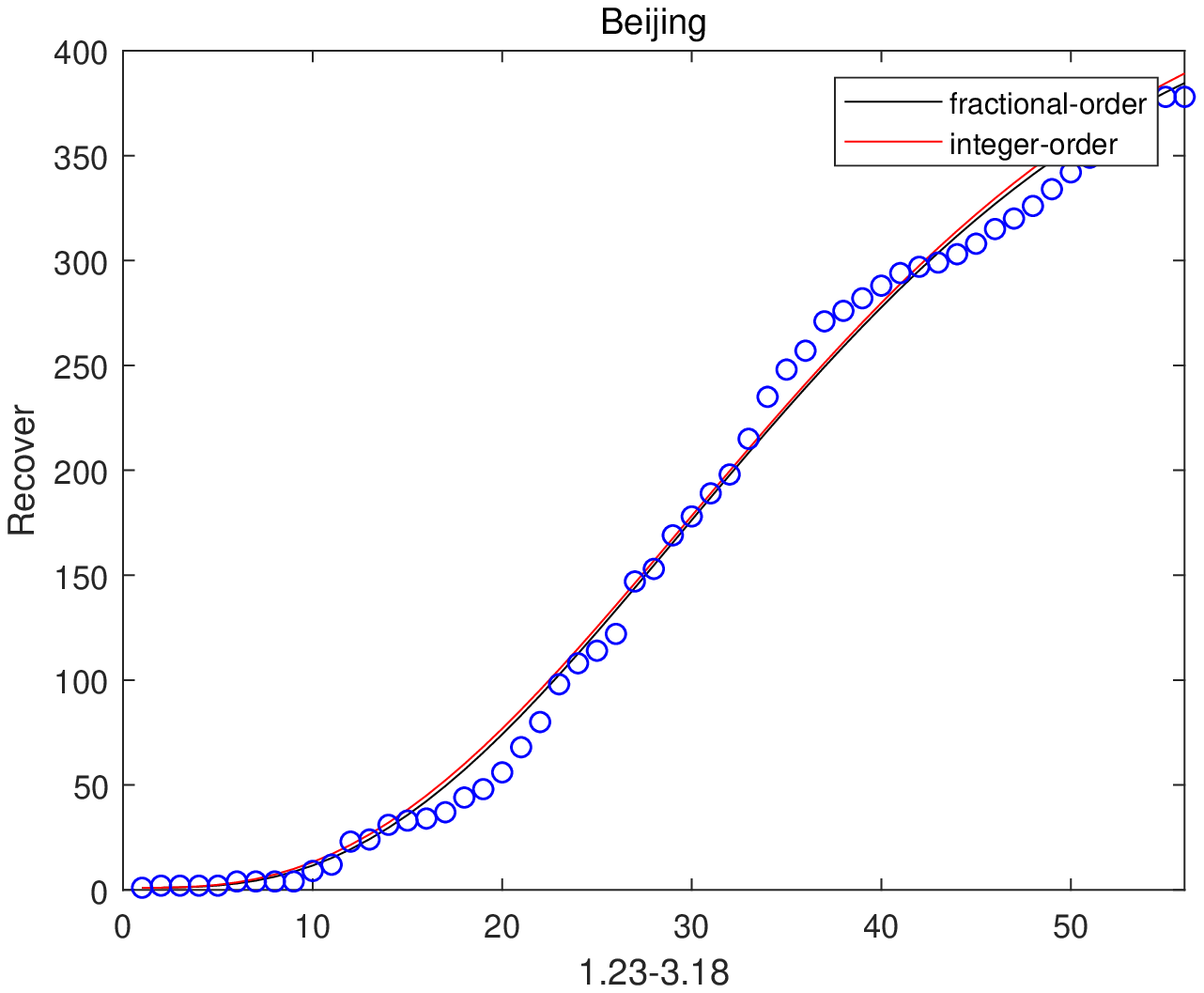}

\caption{Confirmed and Recovered cases of Beijing ($m$=6).}
\label{1}
\end{figure*}

\begin{figure*}
\centering
\includegraphics[width=0.4\linewidth]{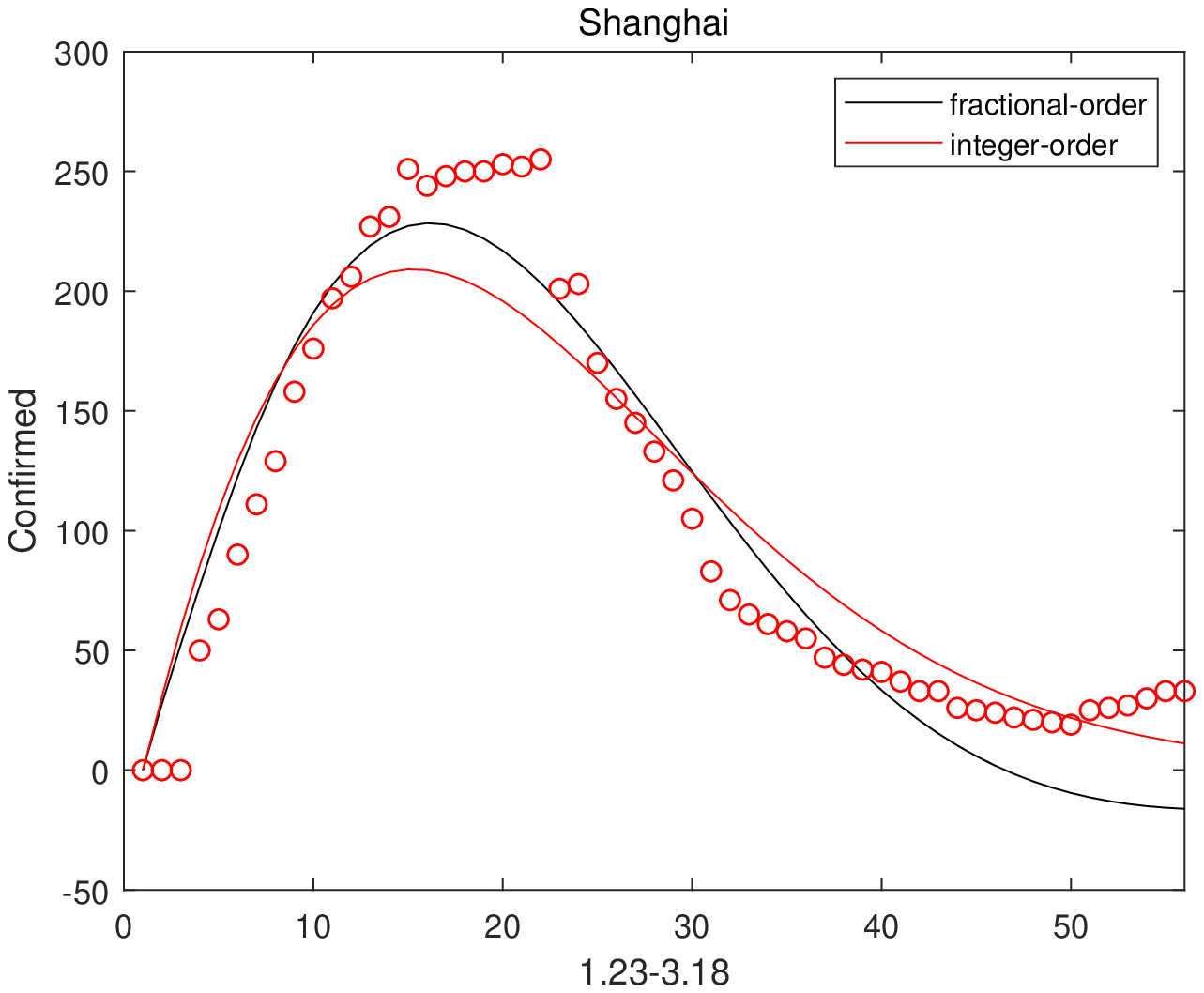}
\includegraphics[width=0.4\linewidth]{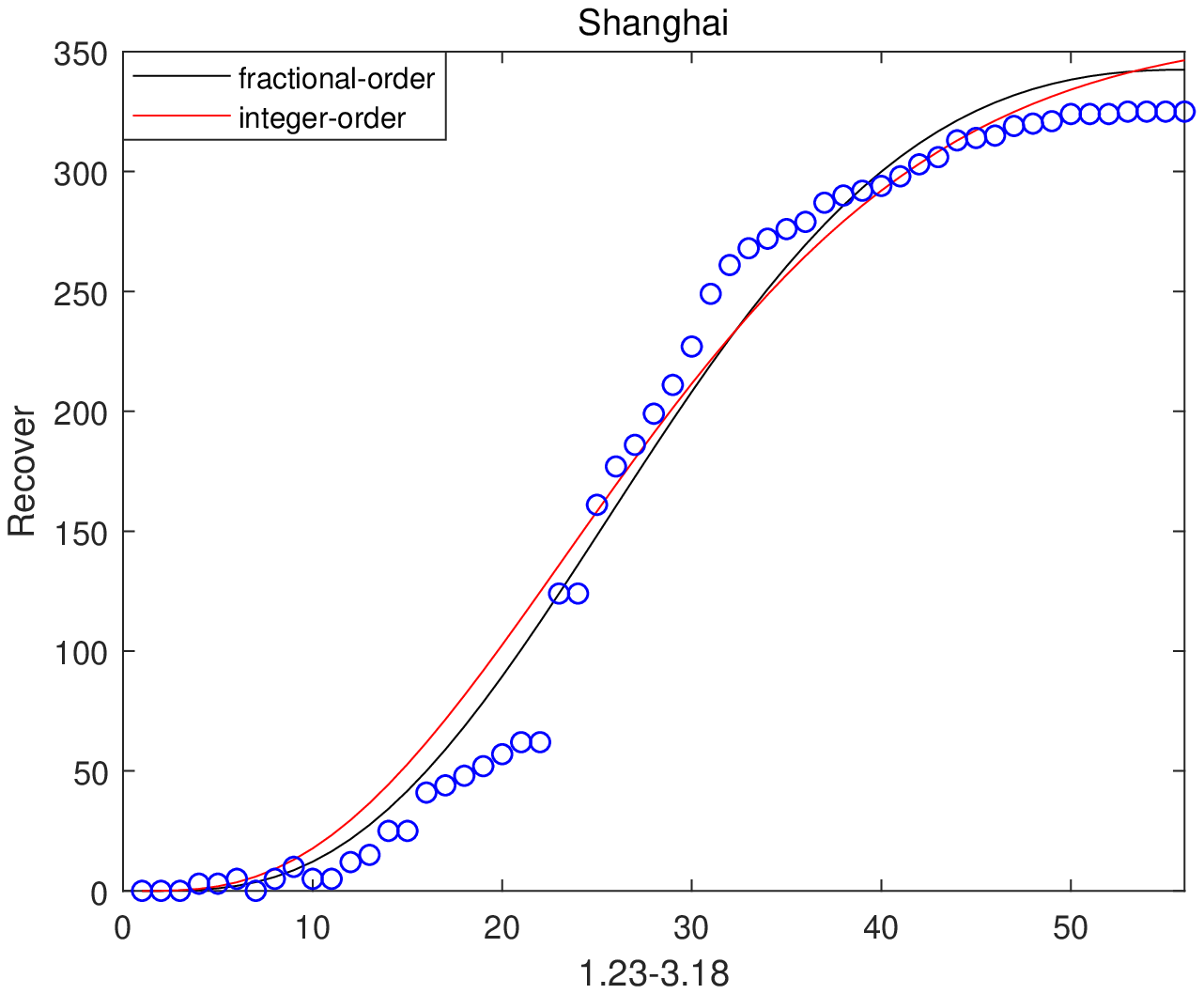}

\caption{Confirmed and Recovered cases of Shanghai ($m$=6).}
\label{2}
\end{figure*}

\begin{figure*}
\centering
\includegraphics[width=0.4\linewidth]{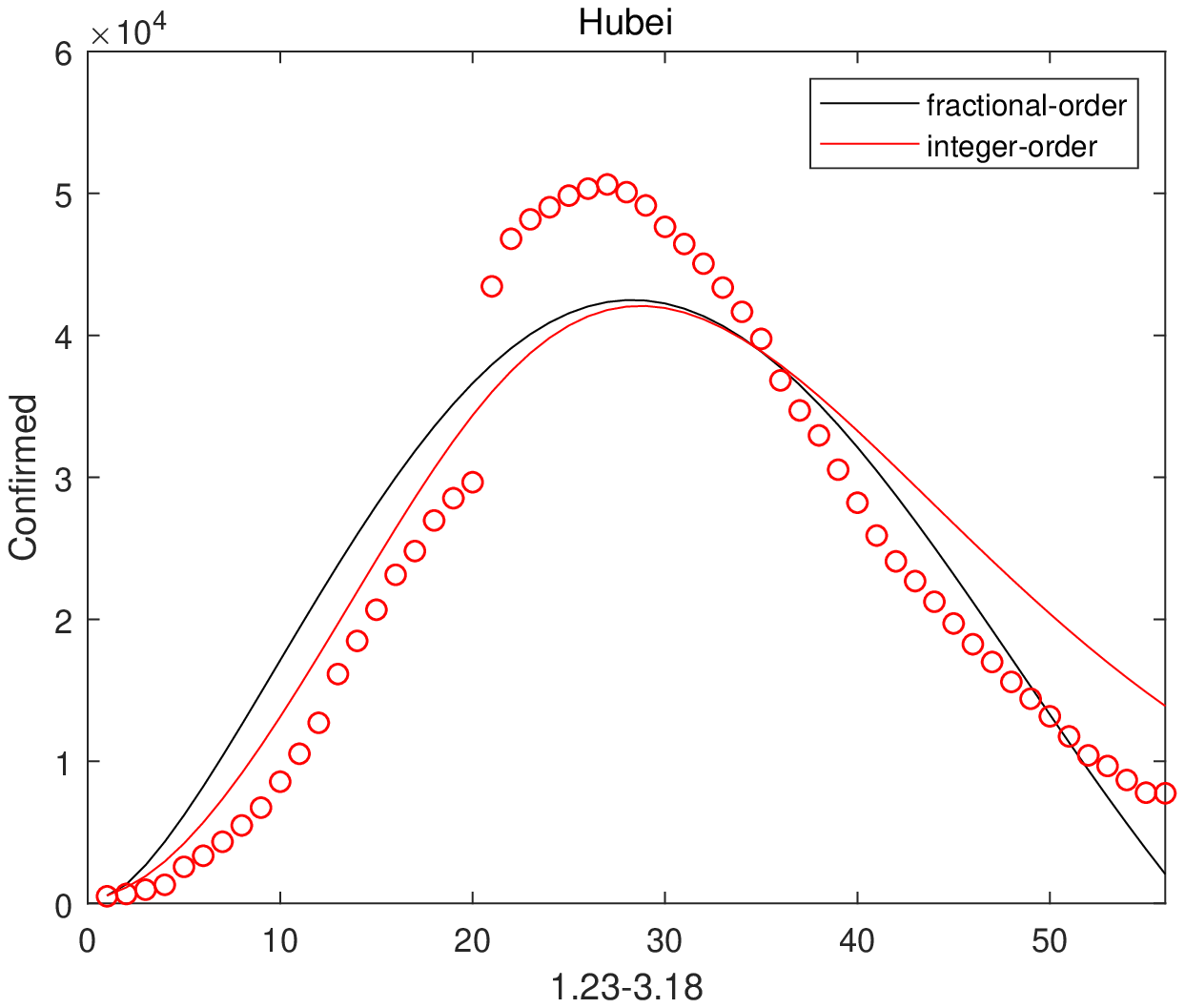}
\includegraphics[width=0.4\linewidth]{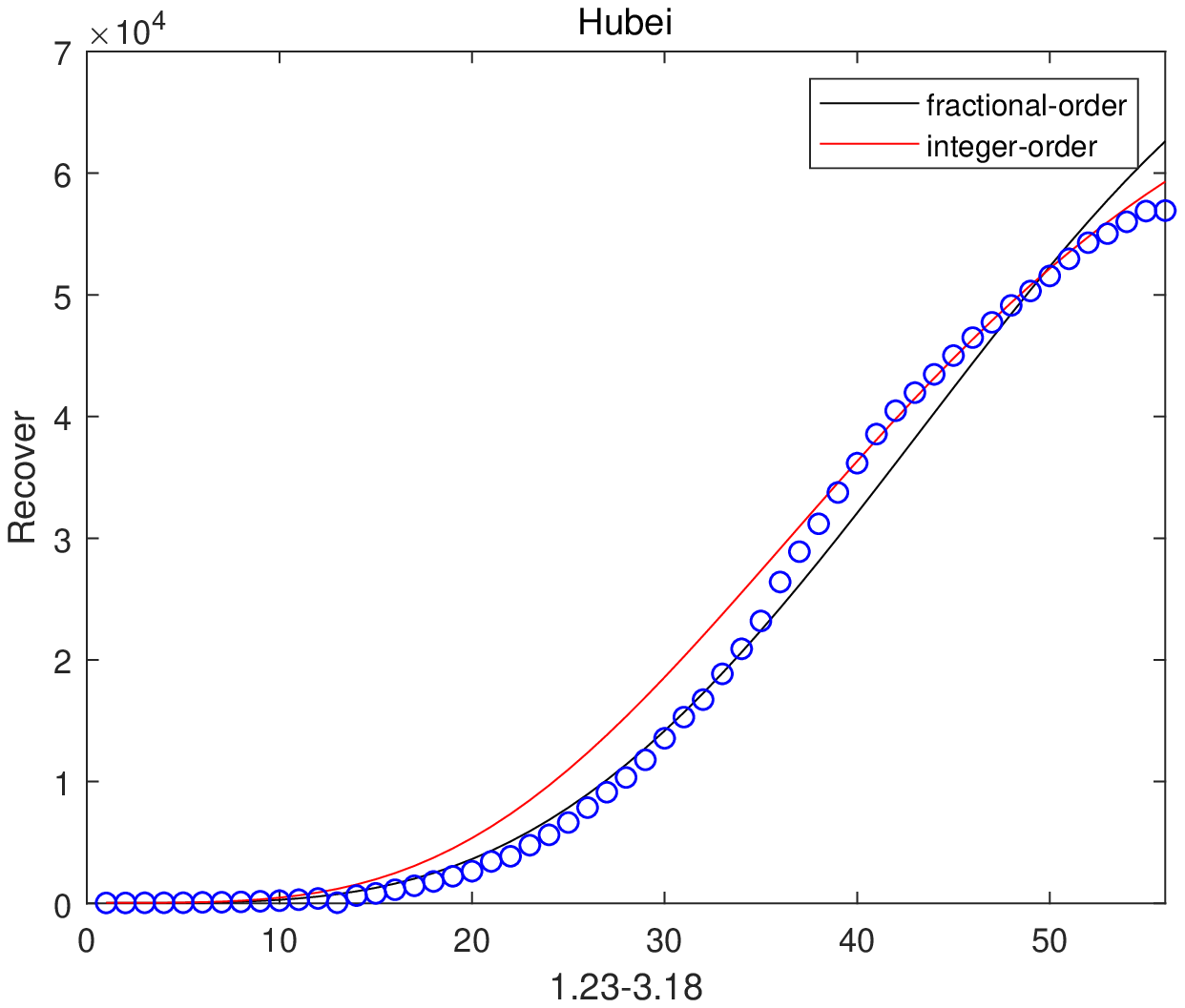}

\caption{Confirmed and Recovered cases of Hubei ($m$=6).}
\label{3}
\end{figure*}

\begin{figure*}
\centering
\includegraphics[width=0.4\linewidth]{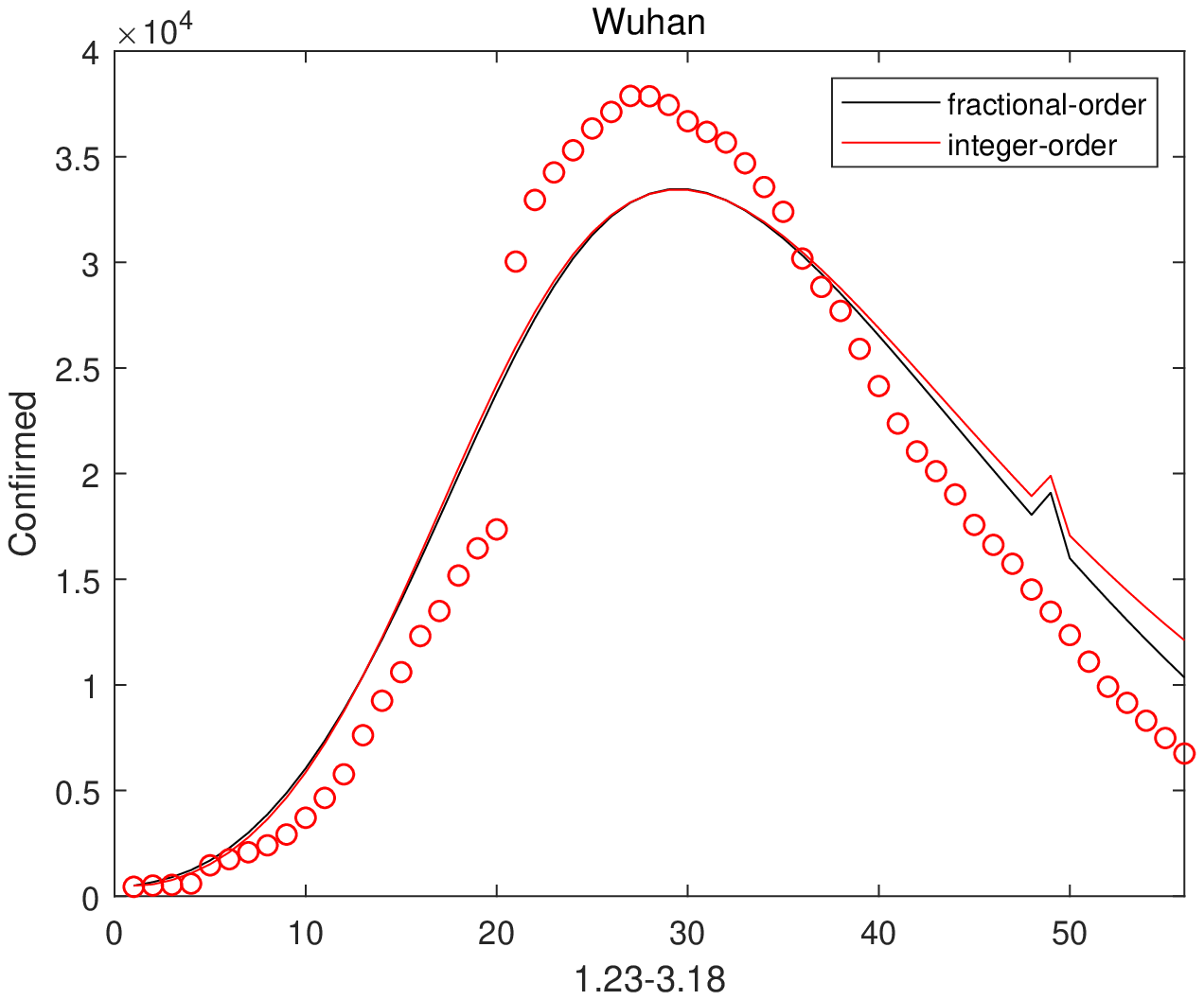}
\includegraphics[width=0.4\linewidth]{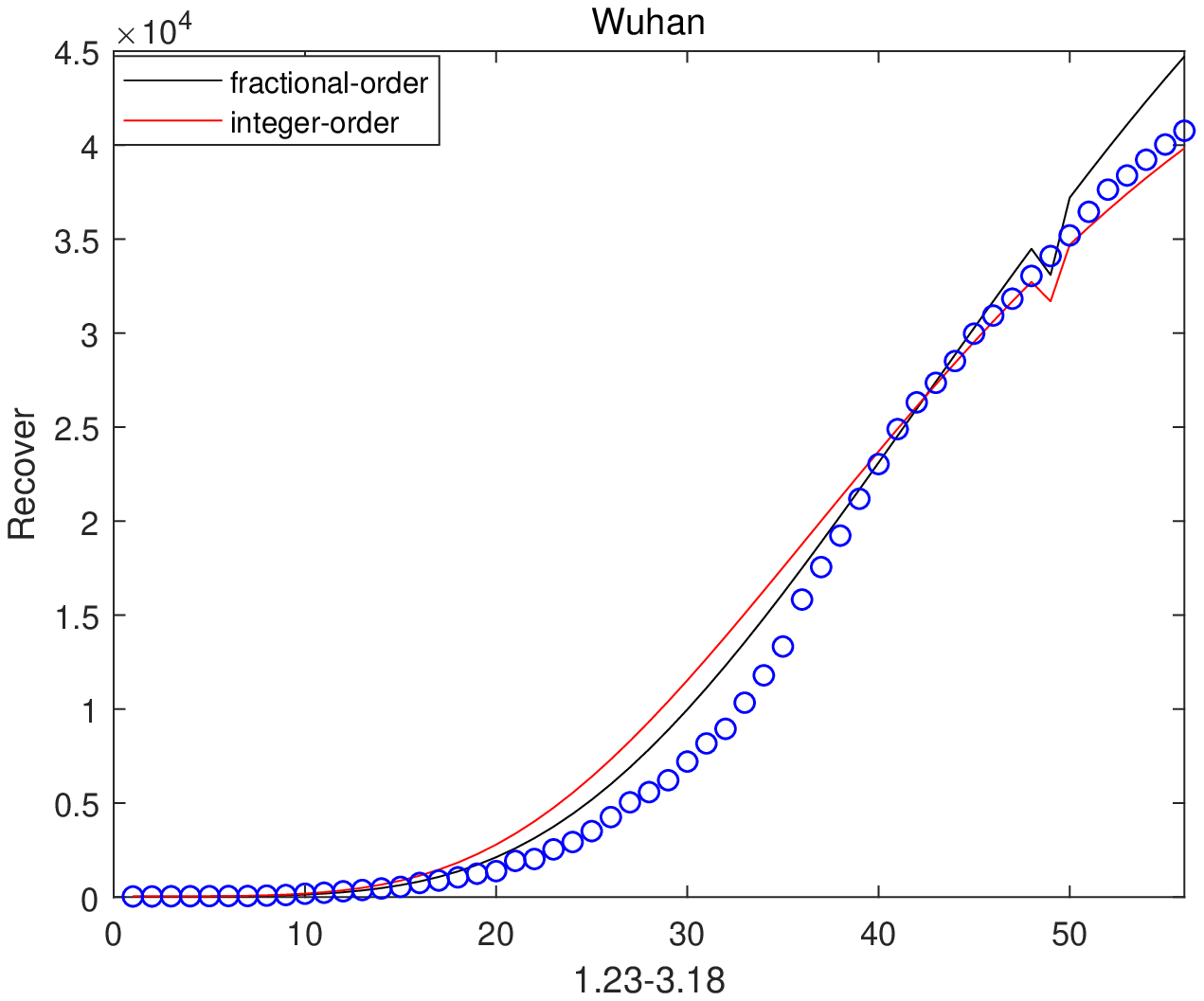}

\caption{Confirmed and Recovered cases of Wuhan ($m$=6).}
\label{4}
\end{figure*}

\begin{figure*}
\centering
\includegraphics[width=0.4\linewidth]{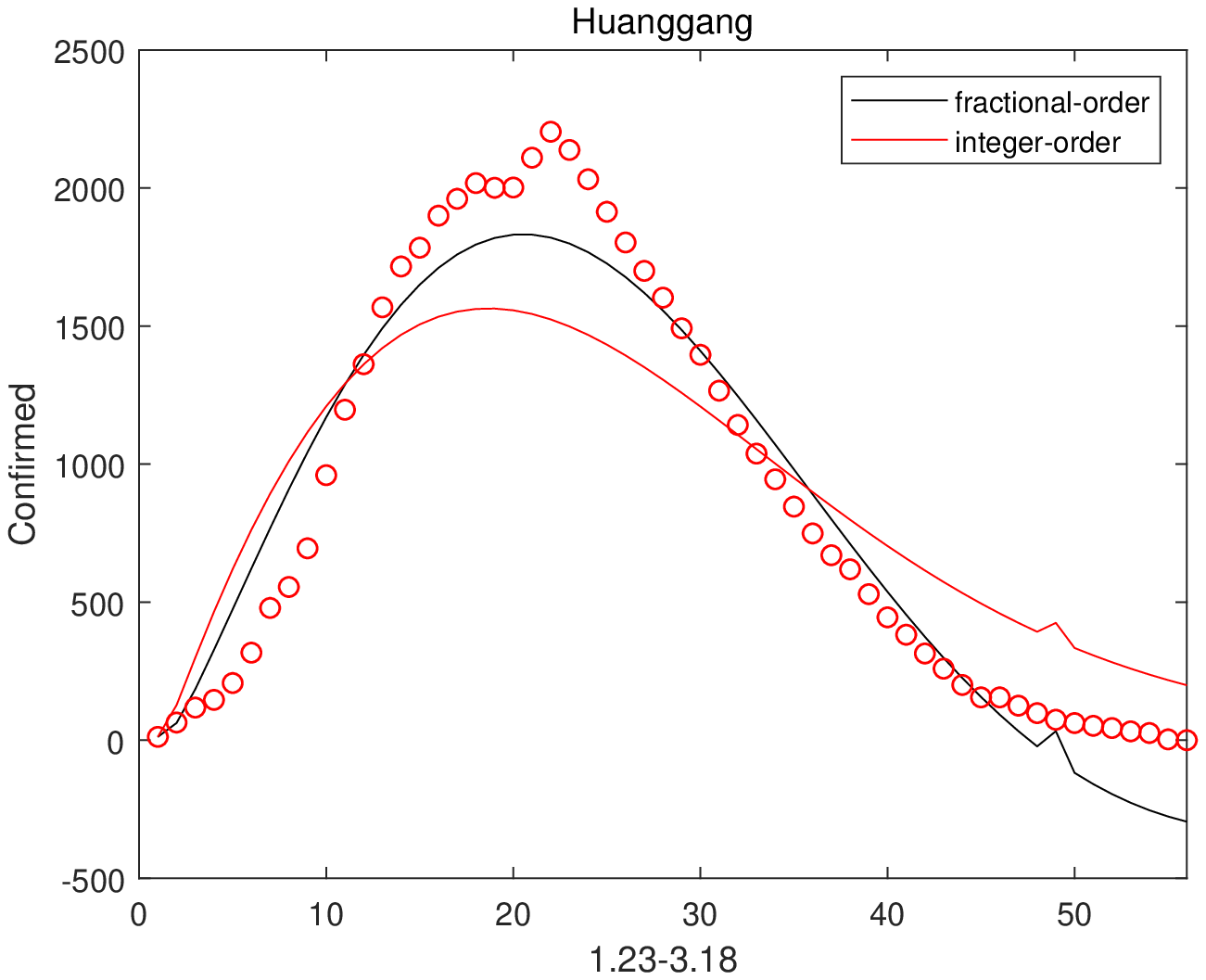}
\includegraphics[width=0.4\linewidth]{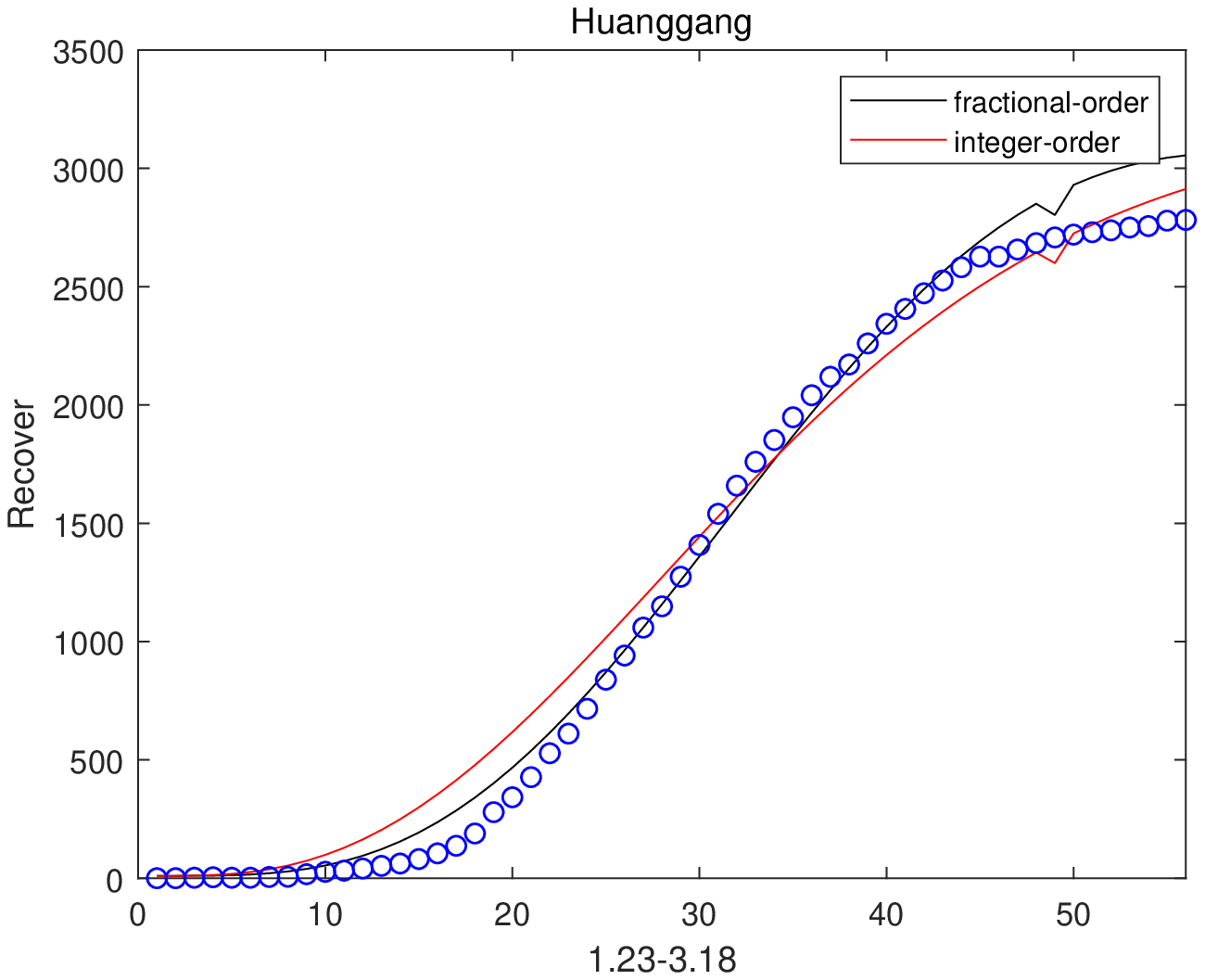}

\caption{Confirmed and Recovered cases of Huanggang ($m$=6).}
\label{5}
\end{figure*}

\begin{figure*}
\centering
\includegraphics[width=0.4\linewidth]{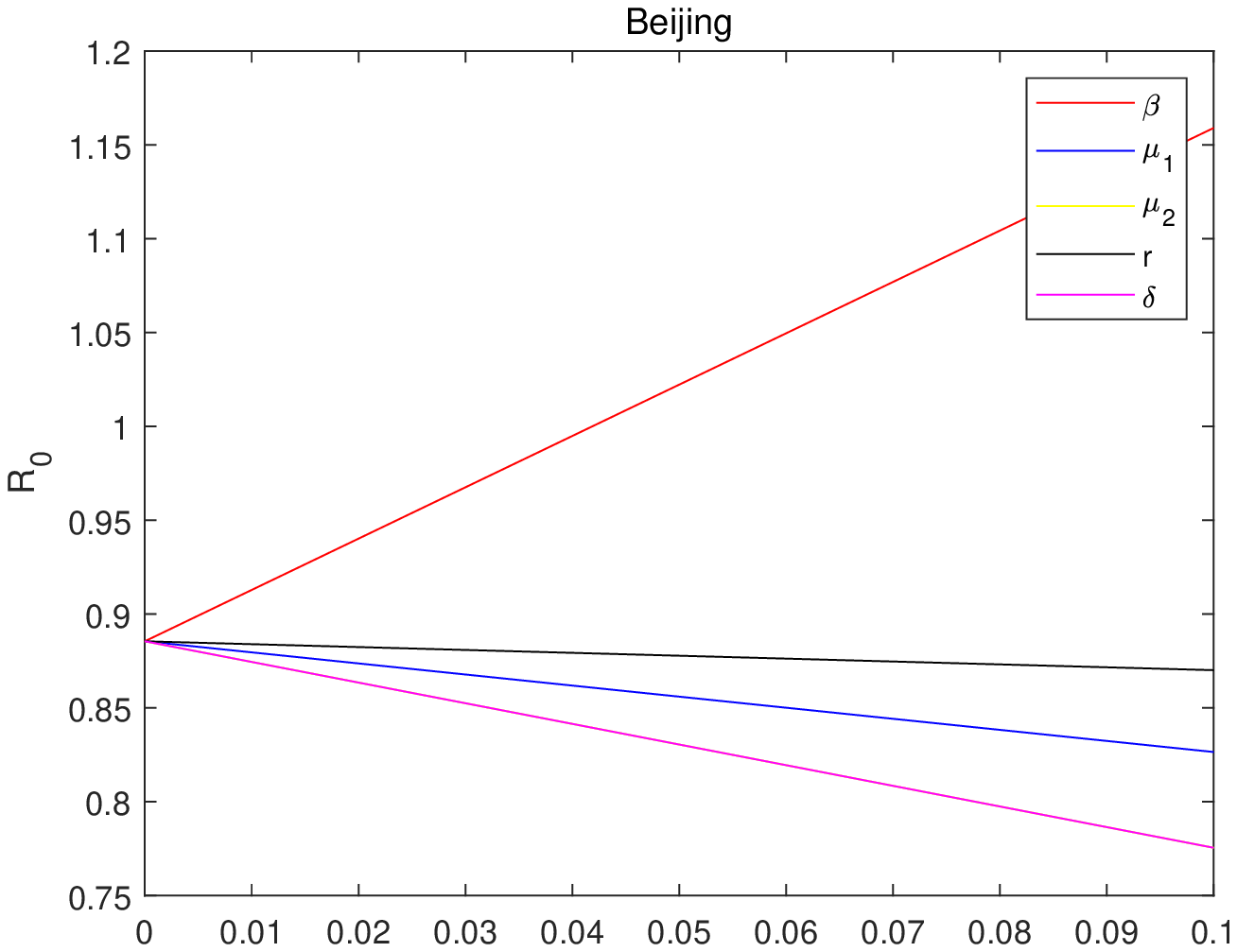}
\includegraphics[width=0.4\linewidth]{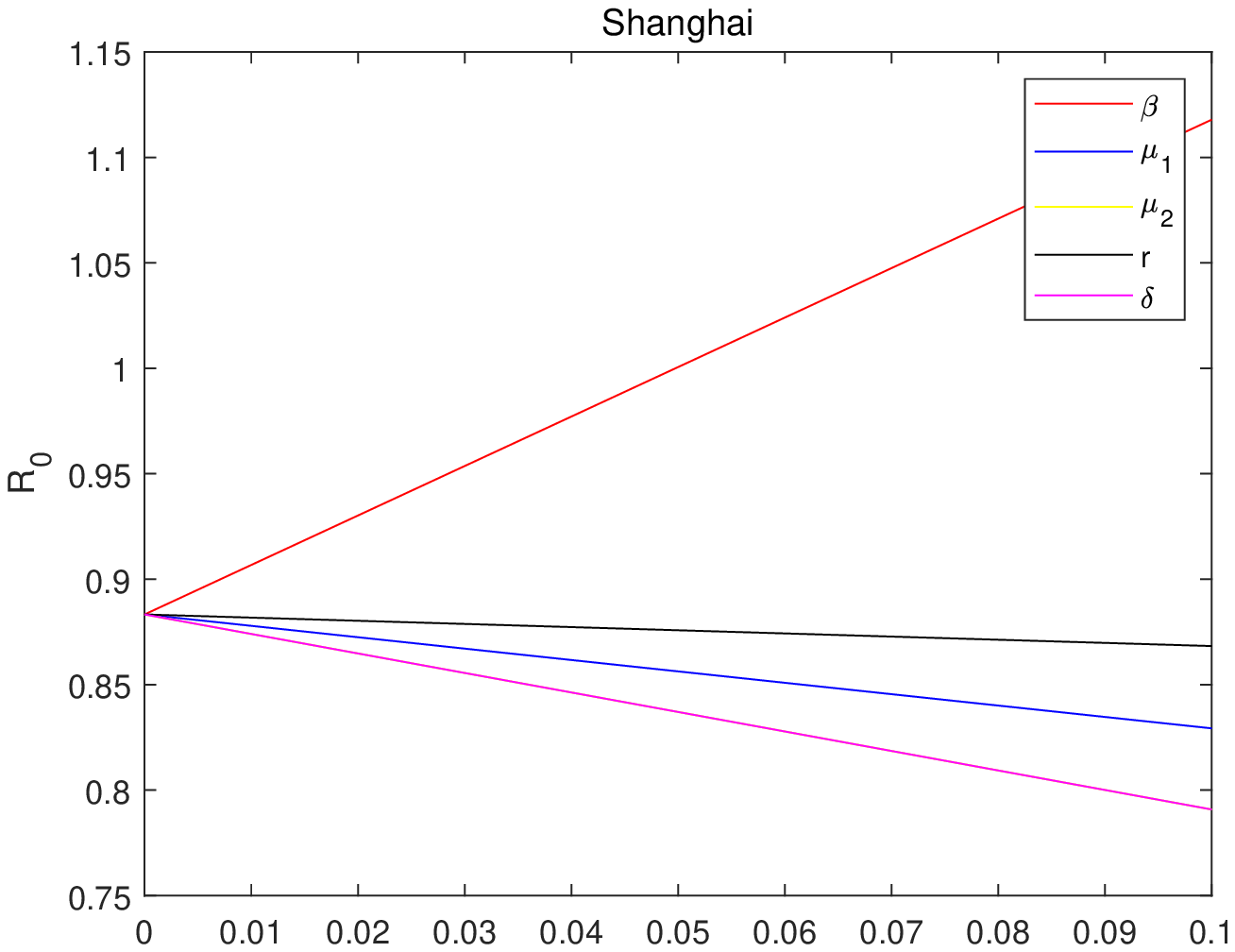}
\includegraphics[width=0.4\linewidth]{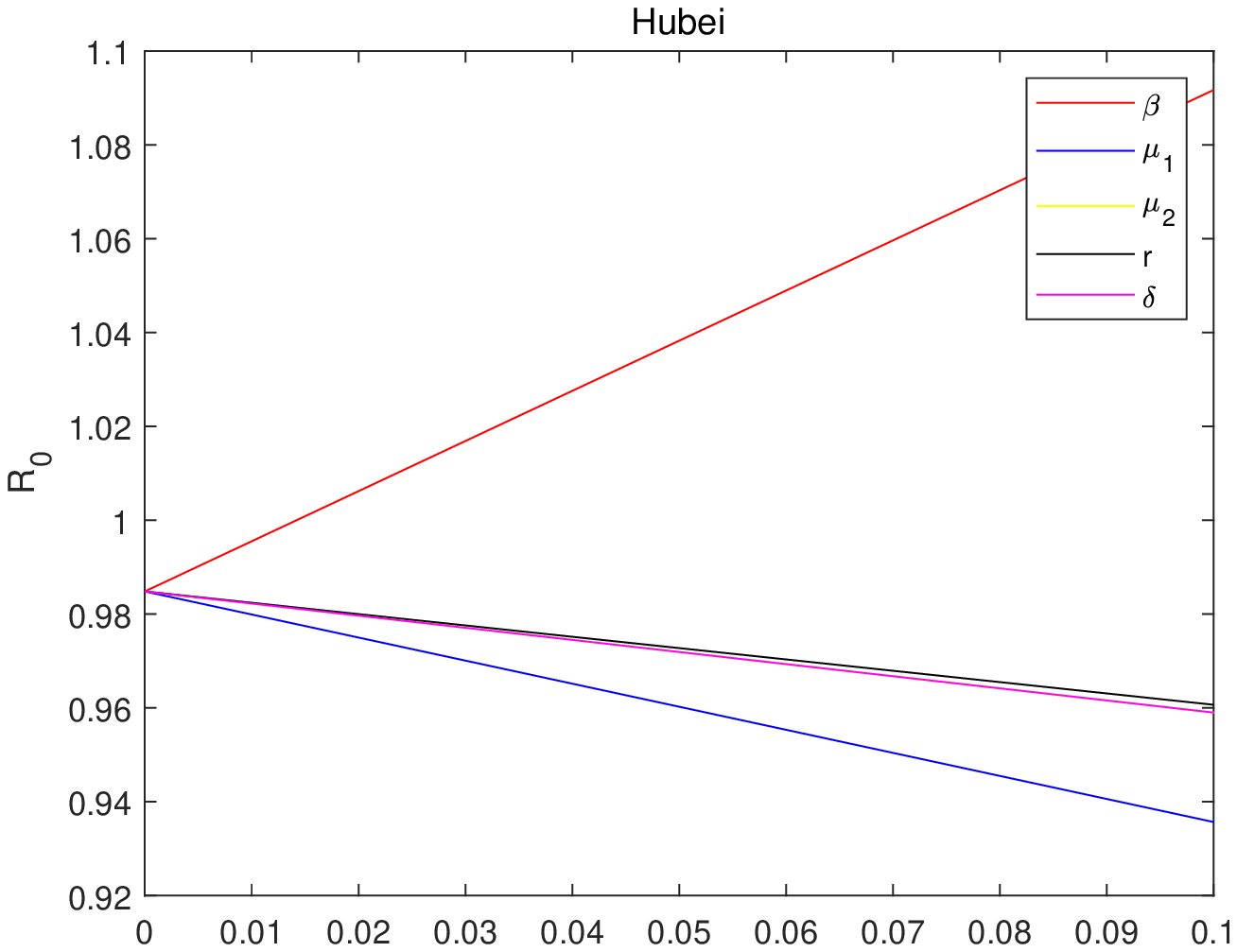}
\includegraphics[width=0.4\linewidth]{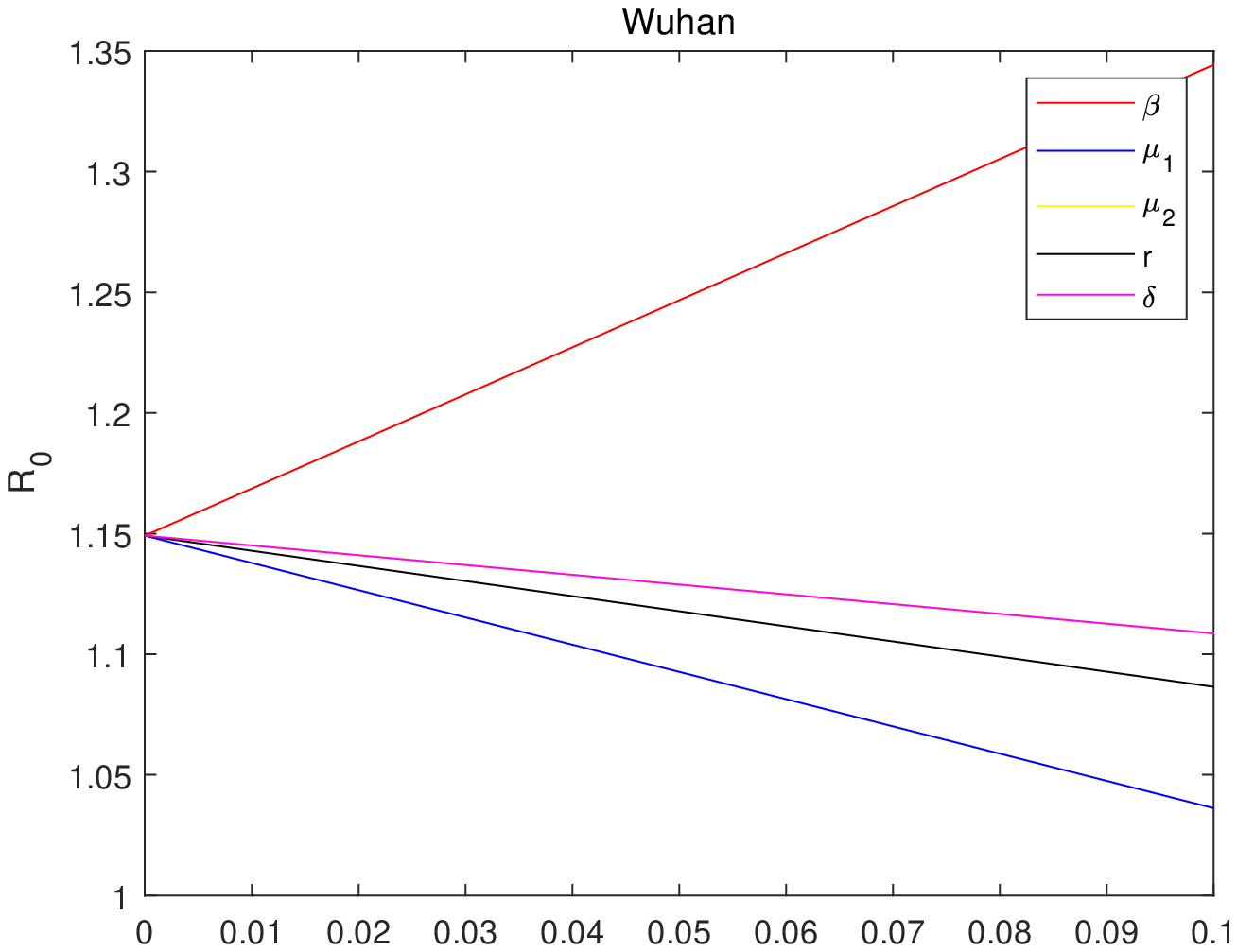}
\includegraphics[width=0.4\linewidth]{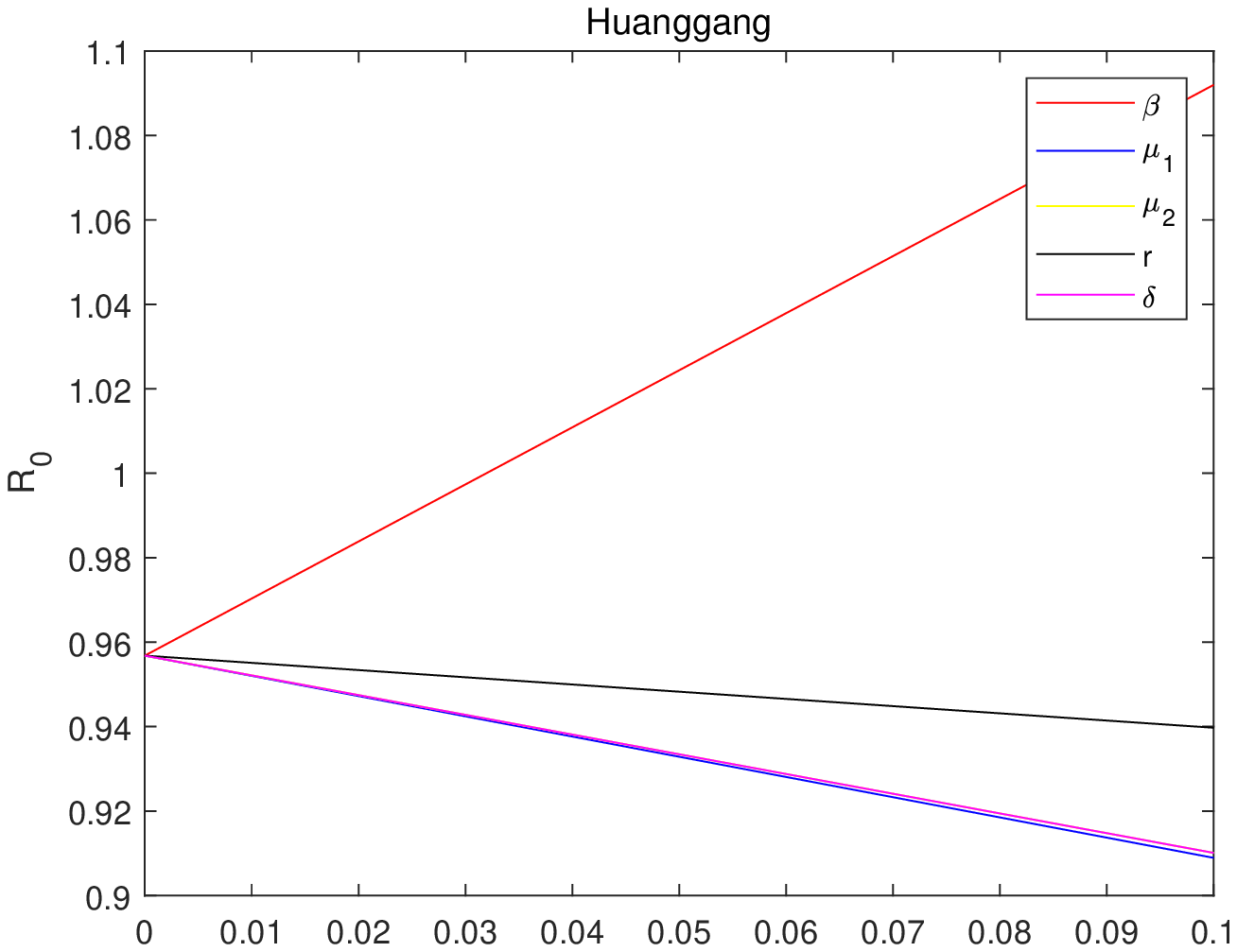}

\caption{The change of the basic reproduction number.}
\label{6}
\end{figure*}

\begin{figure*}
\centering
\includegraphics[width=0.4\linewidth]{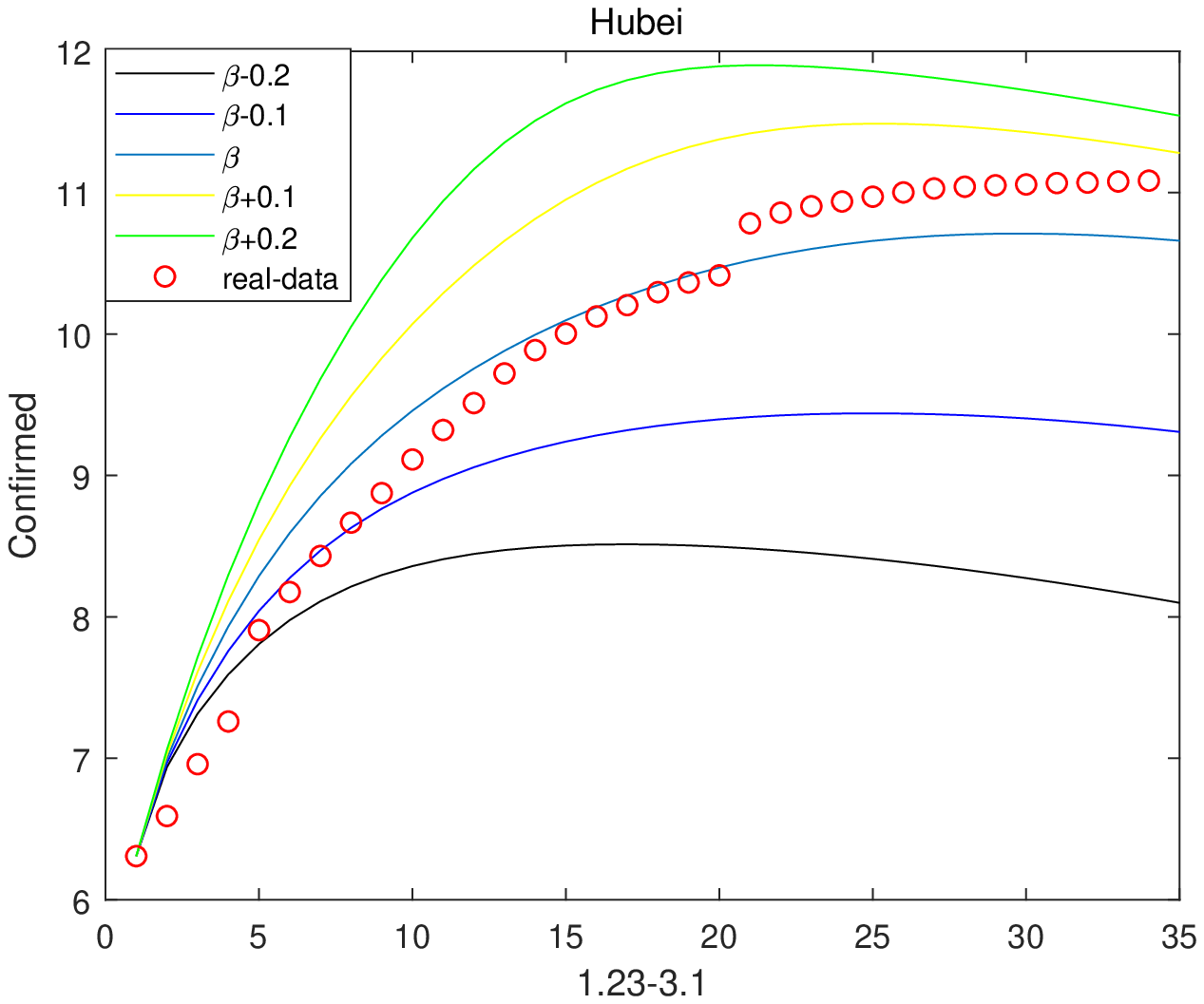}
\includegraphics[width=0.4\linewidth]{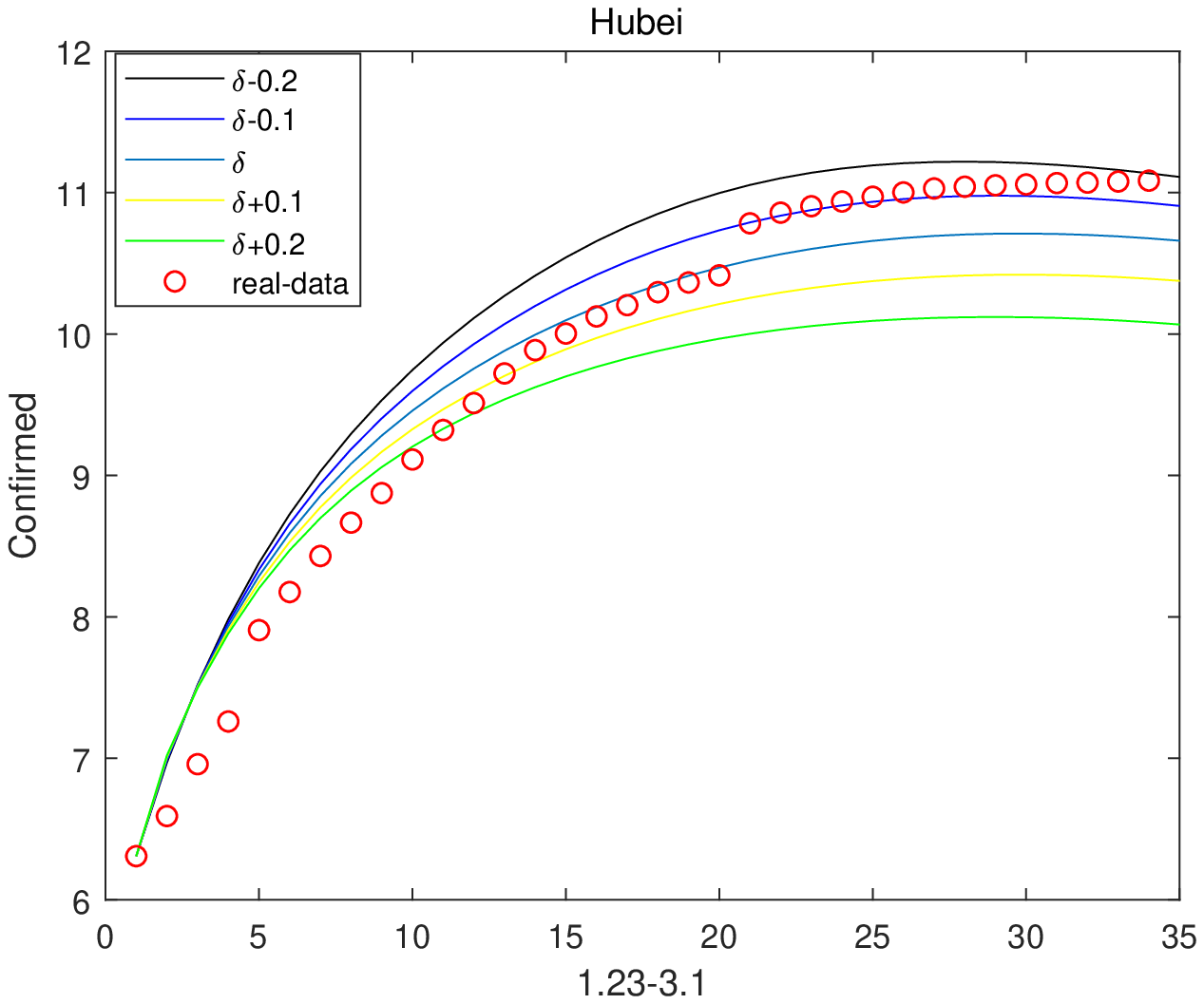}

\caption{Prediction of peak in Hubei province.}
\label{7}
\end{figure*}

From the previous discussion, it can be seen that system (1) is locally stable at the disease-free equilibrium point $E^0$, and it has one unique bounded solution, which may provide a theoretical basis for the coronavirus control. This section presents the numerical stimulation of the coronavirus system (1) by fitting the real-data dating from 21 January to 18 March, 2020, given by NHC every day. System (1) is solved by applying least squares method \cite{u_2020} and predictor-correctors scheme. To evaluate the prediction accuracy, the data can be moved for a fixed number of days, say $m$, prior to March 18. The prediction model is determined upon the data from 23 January up to $18-m$ March, 2020. Then we predict the course of the disease up to 18 March, and the number of omitted days $m$ is equal to the number of prediction days. It is reasonable to assume like in \cite{Peng_2020} from COVID-19 that the cure rates $\lambda(t)$ and the disease-related death rate $\kappa(t)$ vary over time as
\[\lambda(t)=\lambda_0(1-e^{{\lambda_1}t}),~\kappa(t)=\kappa_0e^{-{\kappa_1}t},\]
where $\lambda_0$ and $\kappa_0$ are initial cure and death rate. In the absence of the network (i.e. $k=1$), one can see from Fig.~\ref{1}, Fig.~\ref{2}, Fig.~\ref{3}, Fig.~\ref{4} and Fig.~\ref{5} that the fitting effect of the fraction-order system (1) is better than that of the integer-order system (i.e. $\alpha=1$) both the peak number and time of confirmed individuals. Particularly, the fractional-orders are $\alpha=1.1913$, $\alpha=1.2873$, $\alpha=1.4343$, $\alpha=1.2403$ and $\alpha=0.0332$, respectively. Corresponding the basic reproduction number $R^0_k$ are 0.8855, 0.8833, 0.9848, 1.1491, 0.9568, respectively. Then the sensitivity of the basic reproduction number is presented from Fig.~\ref{6}, which is consistent with Remark 3.2. In addition Fig.~\ref{7}  give a observation iconically on that as the disease transmission coefficient $\beta_{kk}$ and the hospitalization rate of infective individuals $\delta_k$ change, the peak time and final infection size are given (if control measures before January 22 are to continue). The numerical results reveal that strict control of human contact, increased accounting reagents and clinical testing plays a critical role in reducing the number of confirmed case and postponing the occurrence of peak in Hubei province. However, it also can be found from Fig.~\ref{7} that when the disease transmission coefficient $\beta_{kk}$ increases, the number of the infected reaches its peak time earlier, meanwhile, its peak size increases. It can be induced that the measures taken by some countries are feasible and effective.

\begin{figure*}
\centering
 \includegraphics[width=0.4\linewidth]{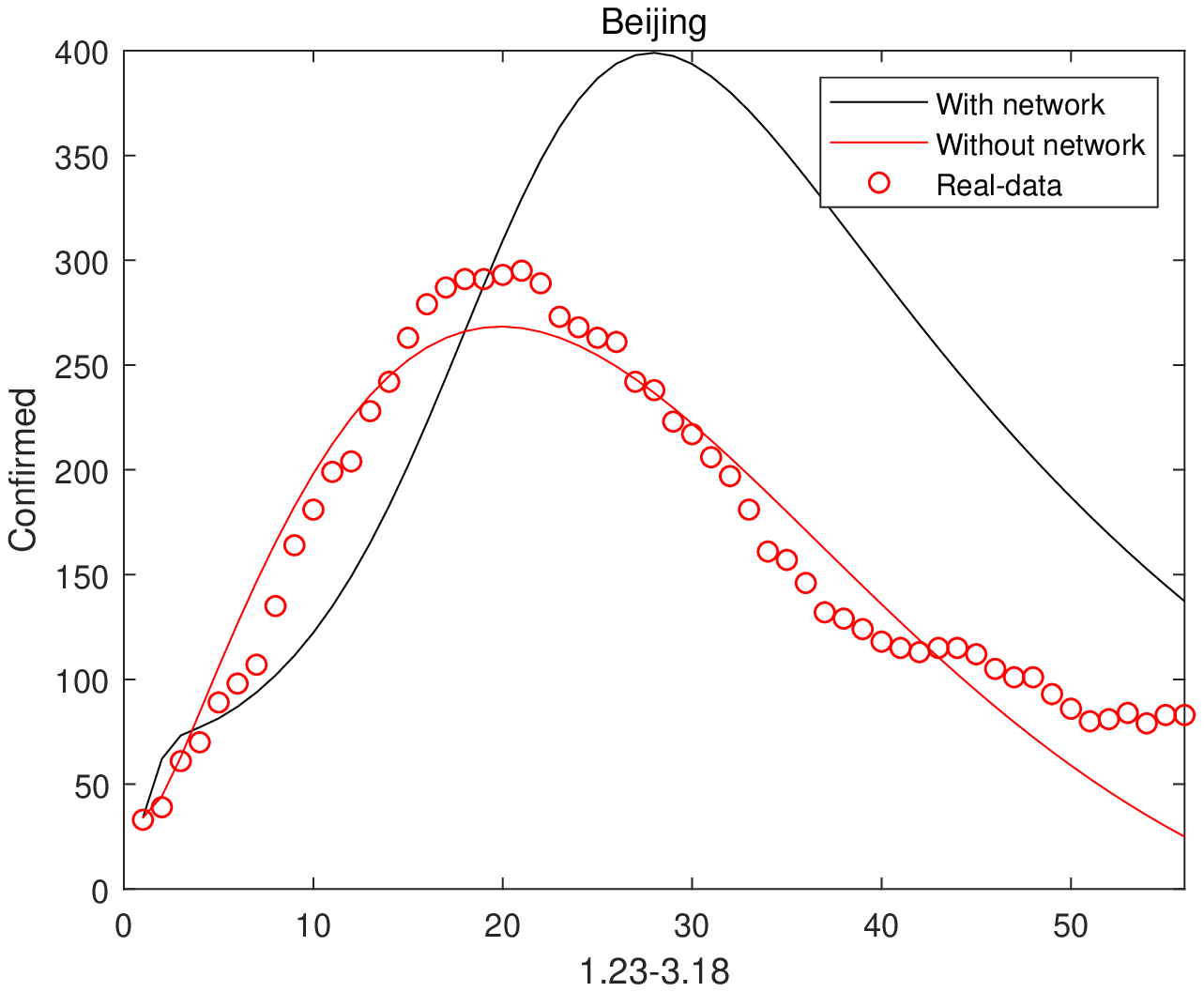}
 \includegraphics[width=0.4\linewidth]{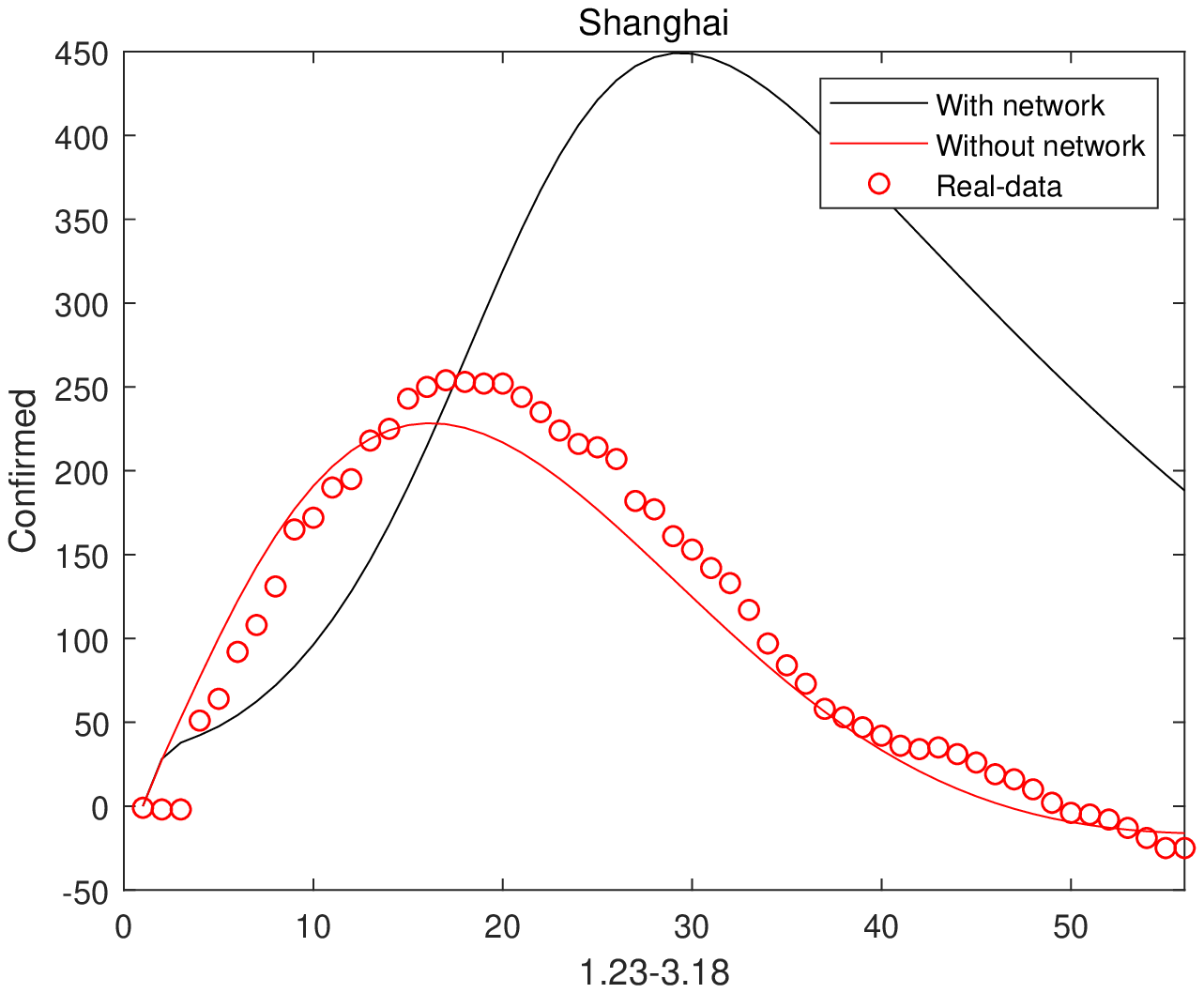}

\caption{Beijing and Shanghai with and without considering intercity network effect.
}
\label{8}
\end{figure*}

\begin{figure*}
\centering
 \includegraphics[width=0.4\linewidth]{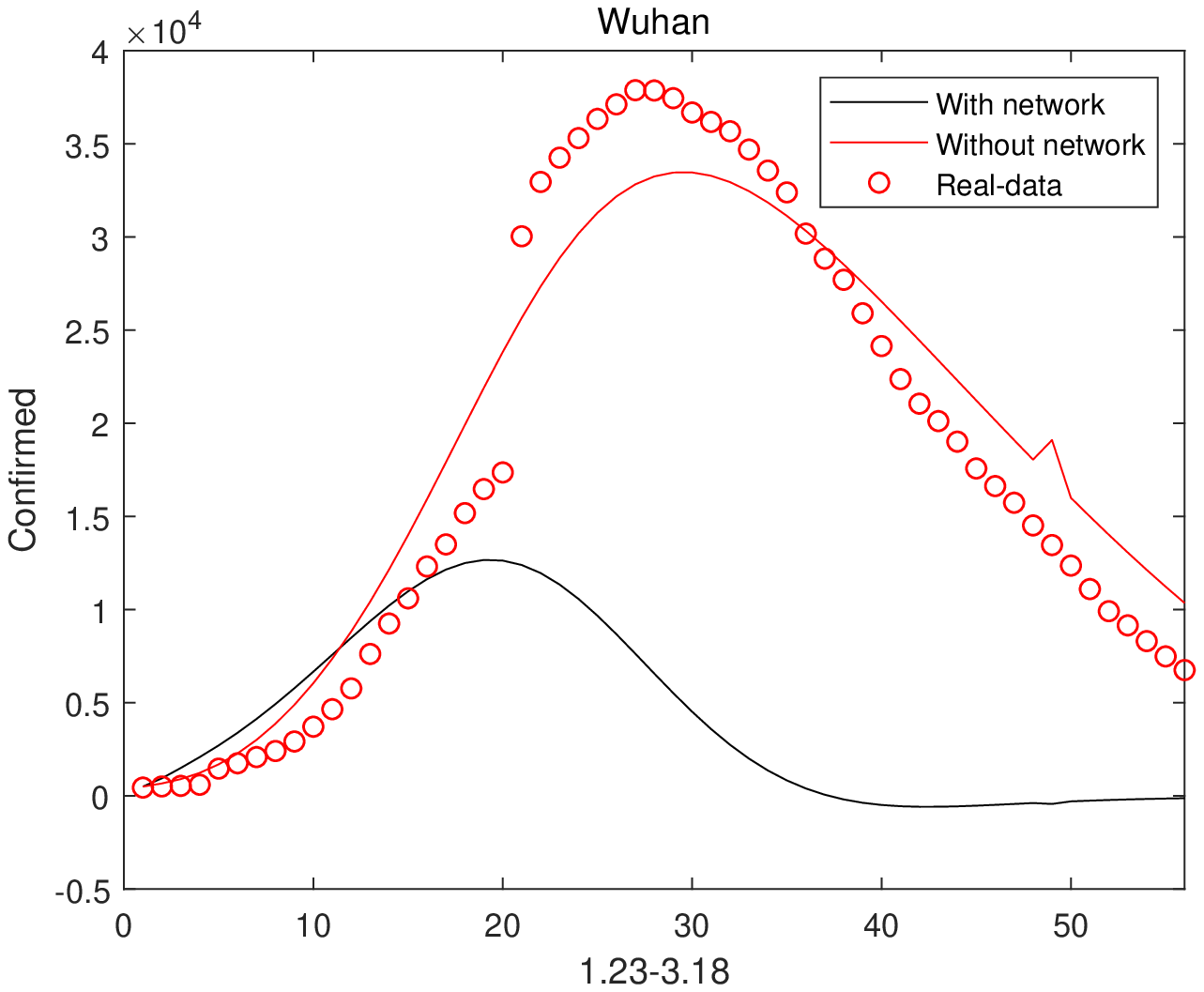}
 \includegraphics[width=0.4\linewidth]{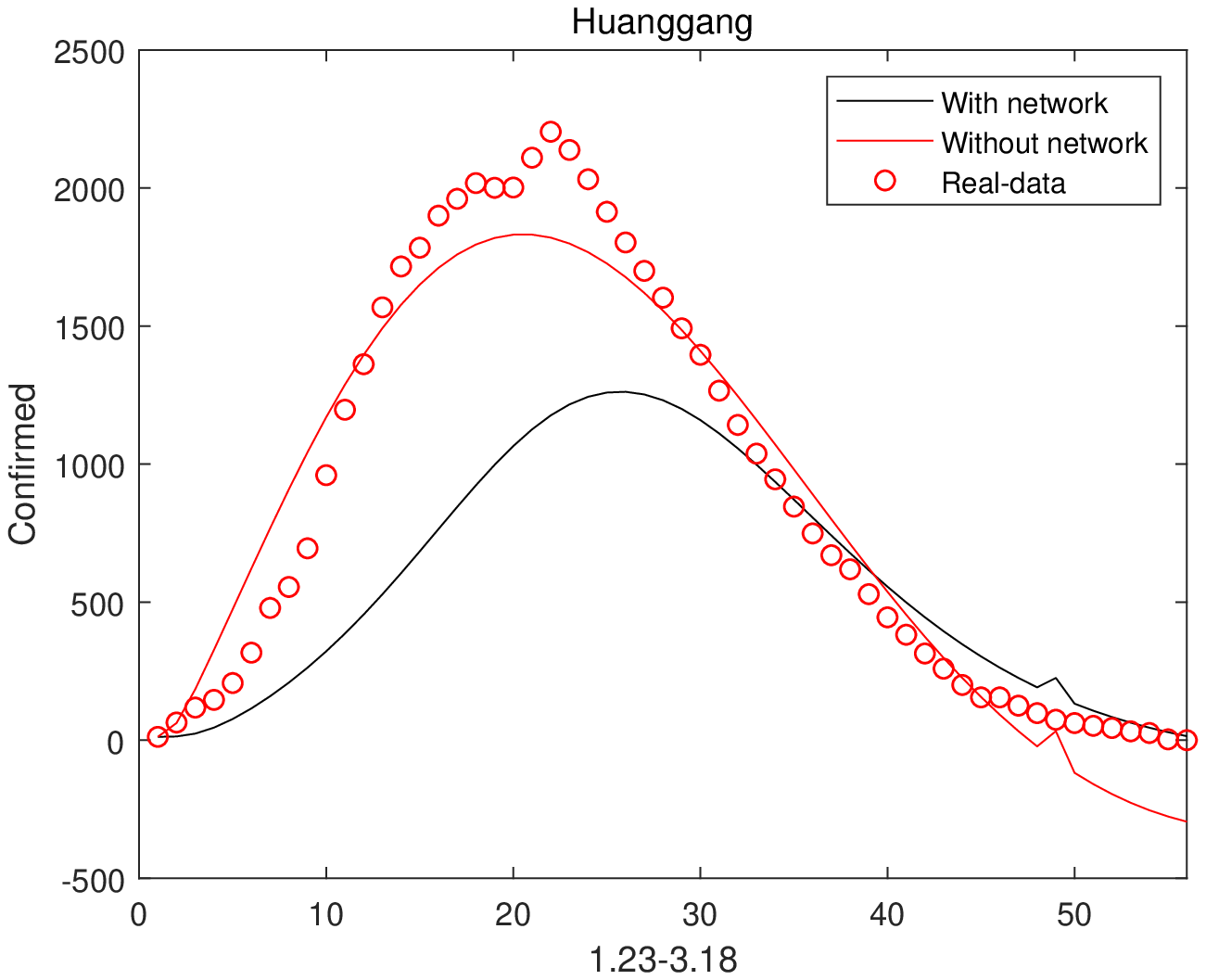}

\caption{Wuhan and Huanggang with and without considering intercity network effect.
}
\label{9}
\end{figure*}

In the case of the inter-city network effect involving the cities in Hubei and their interactions (eg. traffic flow), the precise interactions between cities are known and must be inferred from system (1) by applying least squares method. Then the relationship between Beijing, Shanghai, Wuhan and Huanggang are described in Fig.~\ref{8} and Fig.~\ref{9}. It can be seen that the network-based model is no longer suitable for China, since China imposed city closures on 23 January. Nevertheless, with regard to population mobility in other countries, it can be hypothesized that the network-based model shows a better performance in describing the disease situation which will be the ensuring research focus.

\begin{figure*}
\centering
\includegraphics[width=0.4\linewidth]{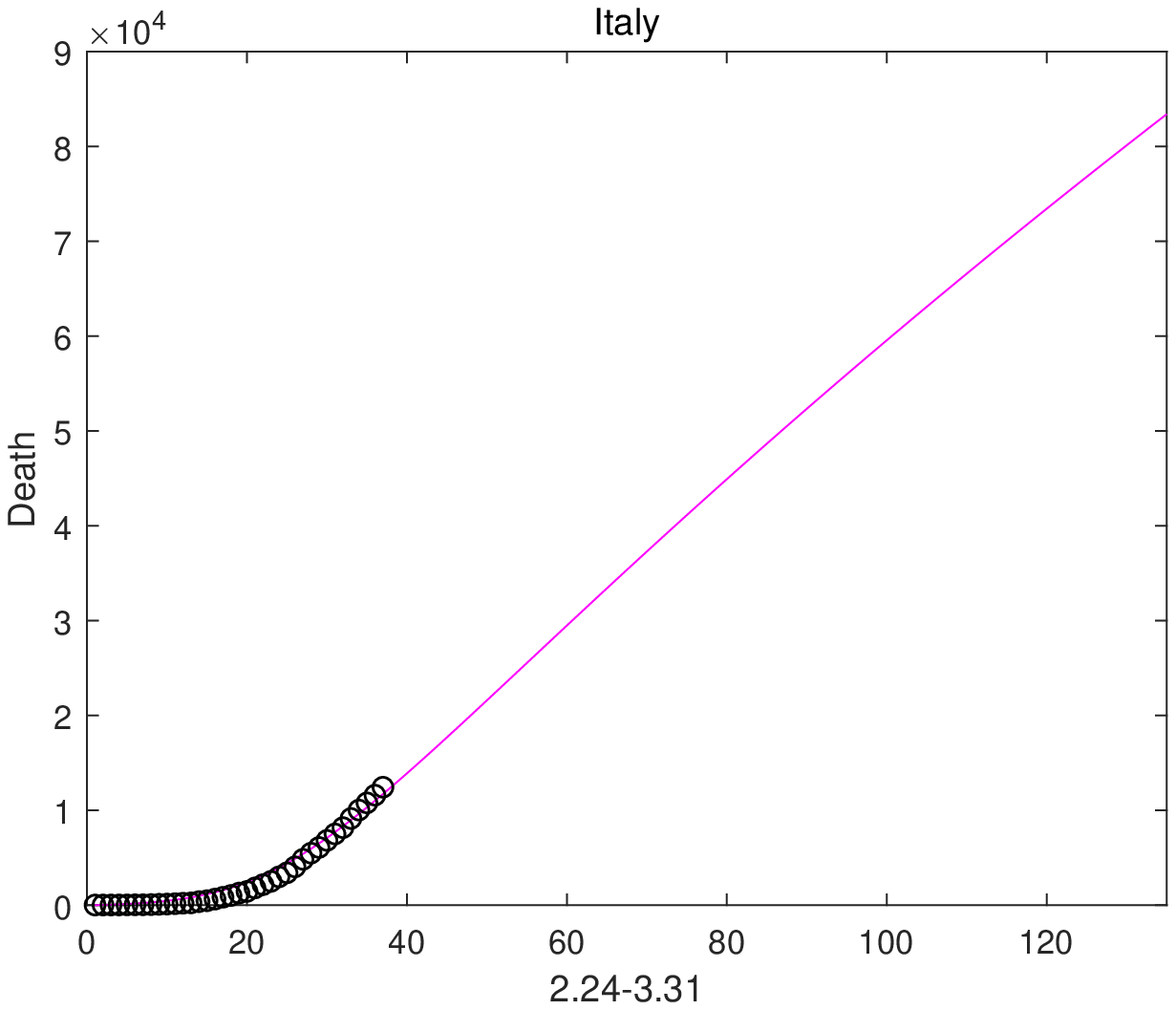}
\includegraphics[width=0.4\linewidth]{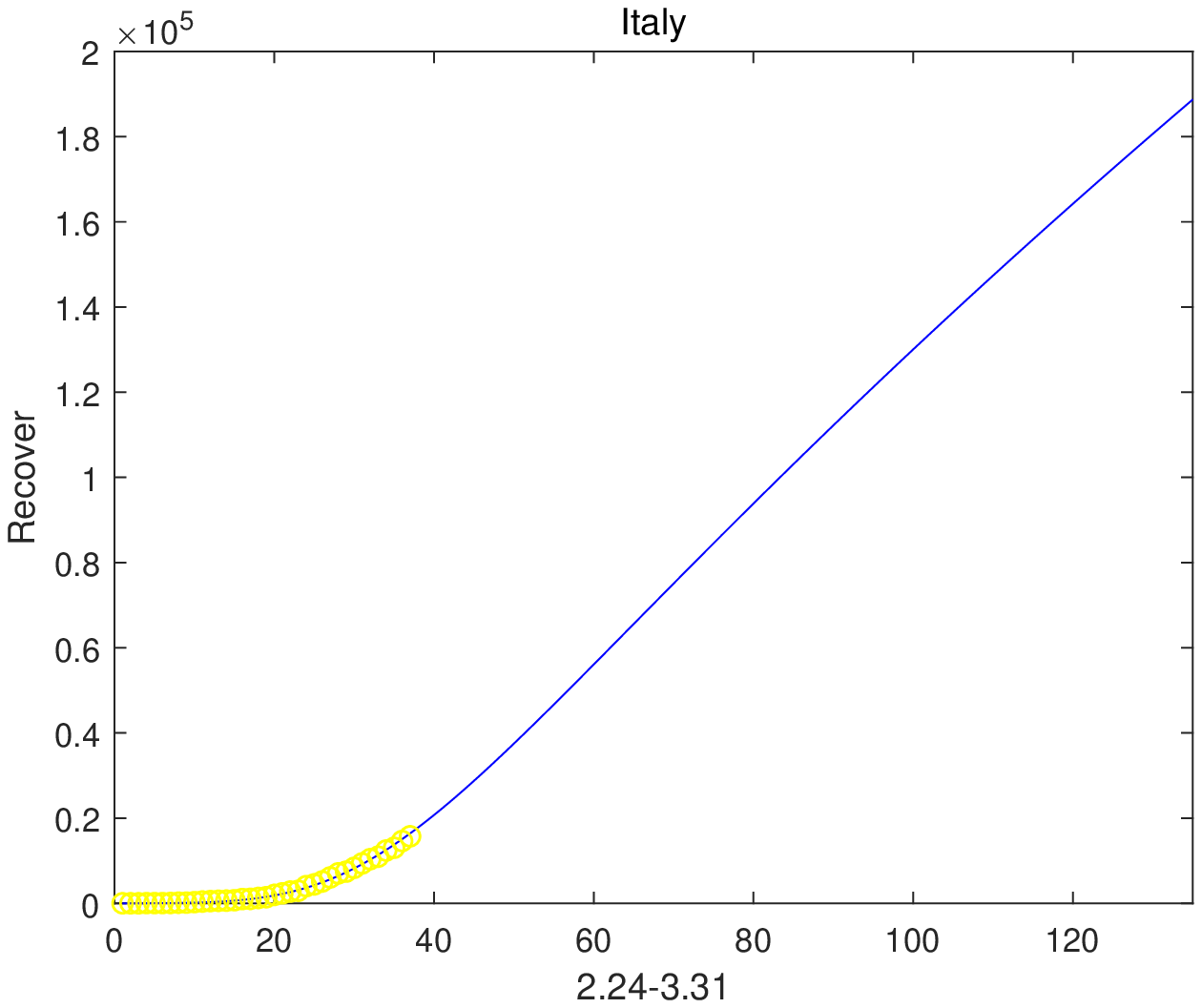}

\caption{Italy outbreak reports from 24 February to 31 March.
}
\label{10}
\end{figure*}

Finally, based on the data of symptoms, hospitalizations, deaths and recoveries in Italy from 24 February to 31 March, the short-term prediction for future outbreaks is realized in Fig.~\ref{10}. Then by calculation, the basic reproduction number $R^0_k=1.3368\times 10^3$ is significantly greater than 1, which suggests that COVID-19 of Italy will not be eliminated in a short term. In addition, it can be seen from Fig.\ref{10} that the number of symptomatic individuals peaked on 30 June and the peak day of the number of hospitalizations occurred on 20 April, without surprise, the number of death and recovery individuals have increased.

\begin{figure*}
\centering
\includegraphics[width=0.4\linewidth]{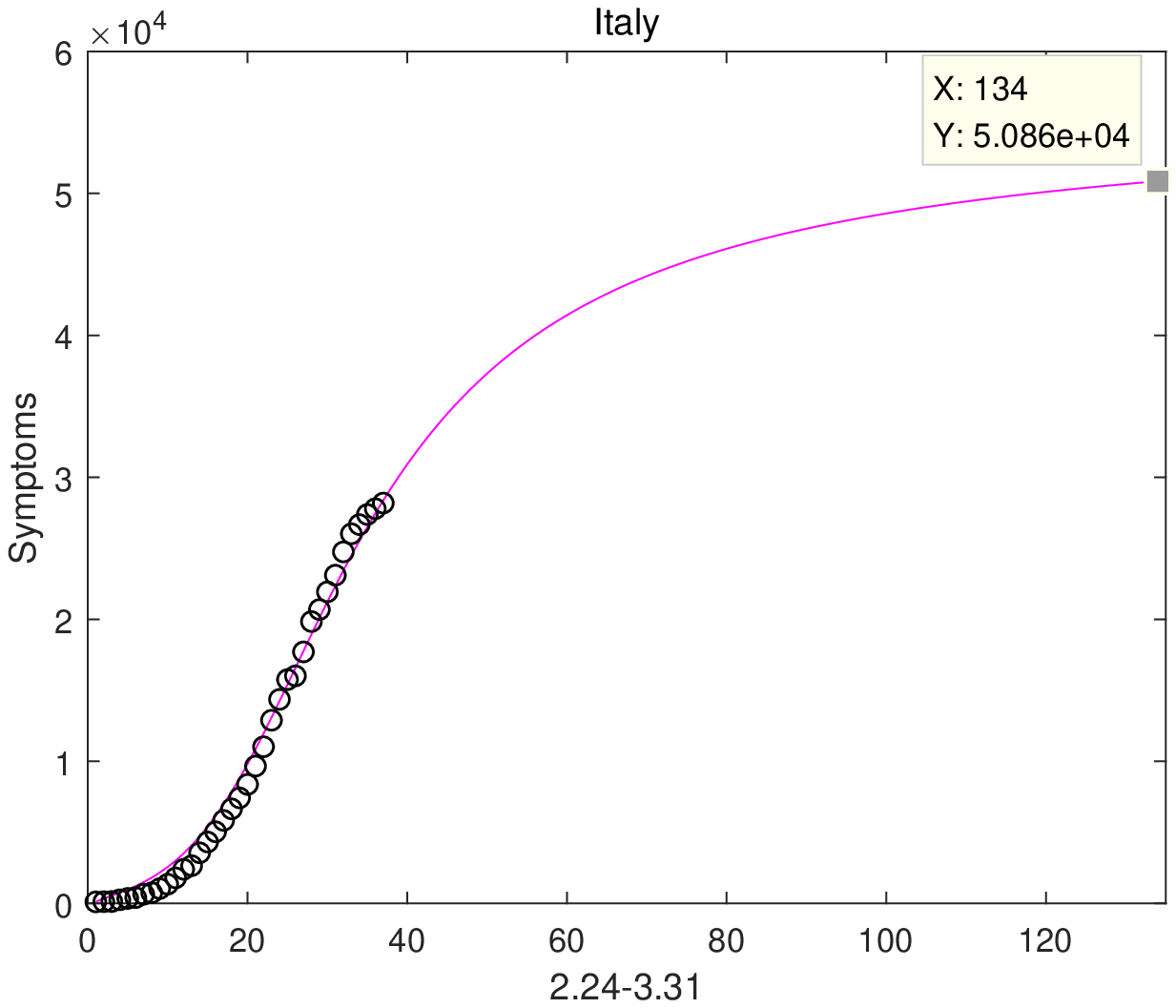}
\includegraphics[width=0.4\linewidth]{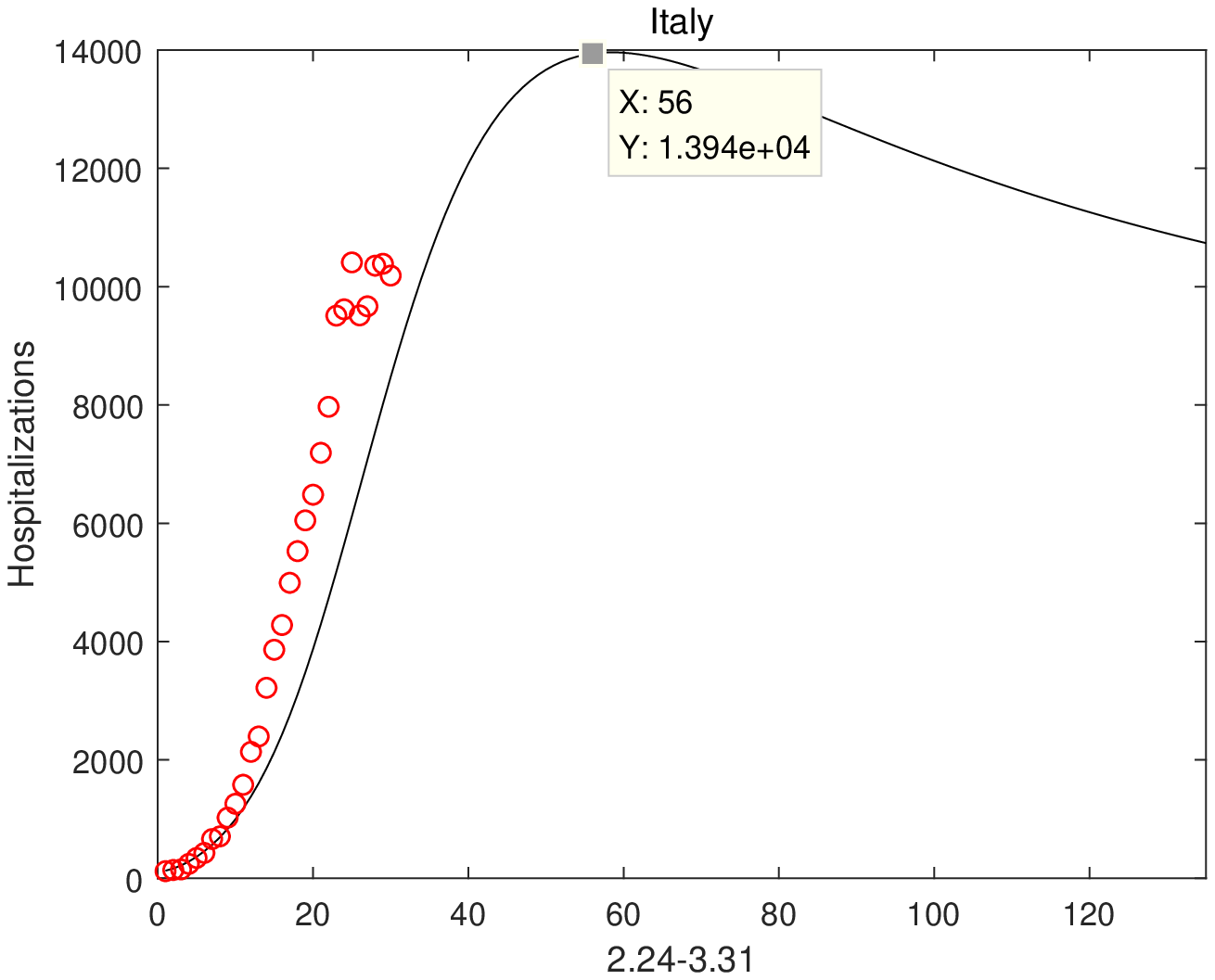}
\end{figure*}

\section{Discussion}
In this paper, incorporating inter-city networked coupling effects for Beijing, Shanghai, Wuhan and Huanggang, a fractional order SEIHDR epidemic model is proposed. By applying least squares method and predictor-correctors scheme, the numerical solution of system (1) is compared with the real-date from 23 January to 18 March, 2020 about Beijing, Shanghai, Hubei, Wuhan and Huanggang. One of the most significant findings to emerge from this investigation is that the fractional-order system has a better fitting effect than the integer-order system. In addition, in the case of the inter-city network effect, the results indicate that the network system may be not a significant case to virus spreading for China. Meanwhile,  system (1) better fits the data from 24 February to 31, March in Italy and the peak number of confirmed individuals in Italy is predicted according to system (1). Moreover, the existence, uniqueness, and positivity solution by the initial-value problem of system (1) are establish. Then the local stability of disease-free equilibrium point are studied by the basic reproduction number $R_0$. And in the absence of network, the sensitivity of $R^k_0$ to the other parameters is analyzed, which provide theoretical basis for disease control.

While this study did not confirm the epidemic model based on inter-city networked effects in China cases due to their strict lock down measures, it did partially substantiate that the fractional-order model based on network may yield more accurate predictions than modeling the epidemic for other countries independently. In spite of its efficiency in predicting the case of COVID-19 in China, this study has inspired several questions in need of further investigation: if the network model can be adapted for predicting the epidemic in other regions; how the factors like medical treatment affect the tendency of virus spreading and so on. It is recommended that further research be undertaken in predicting outbreaks in other countries with networked effects included in the dynamic model.

\section*{Conflict of Interest}

We declare that we have no conflict of interest.

\bibliographystyle{spmpsci}
\bibliography{mybibfiles}

\end{document}